\documentclass[12pt]{amsart}

\usepackage{amssymb}
\usepackage[centertags]{amsmath}

\newtheorem%
{thm}{Theorem}[section]
\newtheorem%
{proposition}[thm]{Proposition}
\newtheorem%
{lemma}[thm]{Lemma}
\newtheorem%
{definition}[thm]{Definition}
\newtheorem%
{corollary}[thm]{Corollary}
\newtheorem%
{conjecture}[thm]{Conjecture} \theoremstyle{definition}

\theoremstyle{remark}

\newcommand{\dontprint}[1]{\relax}

\title
{Consistent interactions and involution}
\author{D.S. Kaparulin, S.L. Lyakhovich and  A.A. Sharapov}
\address{Department of Quantum Field Theory, Tomsk State University, Tomsk 634050, Russia}
\email{dsc@phys.tsu.ru, sll@phys.tsu.ru, sharapov@phys.tsu.ru}

\begin{document}

\maketitle

\begin{abstract}
Starting from the concept of involution of field equations, a
universal method is proposed for constructing consistent
interactions between the fields. The method equally well applies to
the Lagrangian and non-Lagrangian equations and it is explicitly
covariant. No auxiliary fields are introduced. The equations may
have (or have no) gauge symmetry and/or second class constraints in
Hamiltonian formalism, providing the theory admits a Hamiltonian
description. In every case the method identifies all the consistent
interactions.
\end{abstract}

\section{Introduction}
In this paper, a universal method is suggested for solving the
problem of consistent interactions in the general field theories.
The method equally well applies to the Lagrangian or non-Lagrangian
field equations, and it does not break the explicit covariance. The
field equations may have (or have no) gauge symmetry, and/or have
(or don't have) the second class constraints in the corresponding
Hamiltonian form in the variational case - the method provides
identification of all the consistent interactions, in all the
instances.

The proposed method is based on the idea of involutive form of the
field equations. Every regular system of field equations can be
equivalently reformulated in the involutive form. The gauge algebra
of the involutive equations is more rich, in general, then the
algebra of the equivalent non-involutive system. In particular, the
involutive system may have gauge identities even if the theory does
not have any gauge invariance. The consistency is completely
controlled by the stability of the gauge algebra of involutive
system with respect to inclusion of the interaction.

The paper is organized as follows. In the next section, we explain
the problem setting, and discuss previously known methods of its
solution. In Section 3, we explain the  notion of the involutive
form of field equations and preview the basic idea of our method. In
Section 4, we describe the gauge algebra associated to the
involutive system of field equations. Then, we explain the procedure
of perturbative inclusion of interactions in the involutive systems.
In Section 5, we consider the examples of interactions in massive
spin 1 and 2 models, to illustrate the general method. The
Conclusion contains a brief discussion of the paper results. In the
Appendix we derive the relation (\ref{DoF}) which is used in the
paper for counting physical degrees of freedom.

\section{The consistency problem of interactions}
The most common view on the problem of consistent interactions can
be roughly formulated in the following way. The starting point is a
system of linear field equations, or a quadratic Lagrangian, which
is supposed to be covariant with respect to certain global symmetry
group (the most often examples are Poincar\'e, AdS or conformal
groups). The free field model has a certain number of physical
degrees of freedom. Switching on an interaction means inclusion of
nonlinear covariant terms into the free equations. The interaction
is said consistent if and only if the nonlinear theory has the same
number of physical degrees of freedom as the original free model has
had.

If the free field equations and the interacting ones follow from the
least action principle, the Dirac-Bergmann algorithm \cite{Dirac}
will allow one to examine the number of physical degrees of freedom
and to check in this way consistency of the interaction. There is an
extension of this algorithm \cite{LS} which is applicable to general
regular dynamics with not necessarily Lagrangian equations. The
Dirac-Bergmann technique and its extensions, however, break the
explicit covariance. Because of that, these algorithms can help to
examine the consistency of particular interaction a posteriori, but
they are hardly able to serve as a tool for covariant derivation of
the consistent interactions.

The explicitly covariant method is known for solving the interaction
consistency problem for the gauge fields in Lagrangian theories (for
introductory review see \cite{H1}). The basic idea of this method is
that both the action and the gauge symmetry must be simultaneously
deformed by  interaction in such a way that the number of gauge
transformations for the Lagrangian would remain the same after
inclusion of the interaction as it has been in the free theory,
though the symmetry can change. The most systematic form of this
method, being based on the cohomological view of the problem, is
known in the framework of the BV-BRST formalism \cite{BH1},
\cite{H1}. In its turn, the BRST approach to the consistency problem
is based on the general theory of local BRST cohomology \cite{BBH}.
Many gauge field theories are known where this method either has
delivered the complete solution to the consistency problem, or
established a no-go theorem for the interactions. We mention several
examples of such results: establishing all the consistent
interactions for $p$-form fields \cite{PF}, and the no-go theorem
for graviton-graviton interactions \cite{MGT}. There are numerous
results of a similar type obtained in various models by various
versions of this method during the two recent decades. Among the
most recent ones we mention the work \cite{HGR}, where all cubic
electromagnetic interactions are found by this method for the
higher-spin fermionic fields in Minkowski space, and it is proven
that the minimal couplings are not admissible.

 The above mentioned method is so popular because of the explicit covariance and algebraic elegance.
 This method, however, is unable to provide a solution for the consistency interaction problem for any field theory.
 The matter is that gauge invariance is not the only cause that makes ``nonphysical'' some of degrees of freedom.
 For example, the number of physical degrees of freedom is less than the number of fields for the second class constrained models,
 though there is no gauge symmetry.
 That is why, even in gauge invariant systems, the non-violation
 of gauge symmetry by the interaction does not necessarily mean the consistency of interaction.
 With this regard, it is relevant to mention the idea of introducing St\"uckelberg fields to gauge the massive models,
 and then apply the usual method to introduce the gauge invariant interactions.
 Many works apply the St\"uckelberg gauge symmetry idea to find the consistent
 interactions or examine consistency for various massive fields. Among the examples of this type
 we can mention the series of recent works on massive gravity \cite{MaGra1}, \cite{MaGra2}, \cite{MaGra3}
 (further references can be found in these articles, and also in \cite{Zi}).
 Inclusion of the St\"uckelberg fields is rather an art than a science at the moment.
 No general prescription is known of doing that in such a way that could ensure the consistency
 of interaction just as a consequence of its consistency with the St\"uckelberg symmetry.
 To examine consistency of interactions in the St\"uckelberg gauged model,
 a non-covariant Dirac-Bergmann constrained analysis still remains inevitable in many cases,
 see for example \cite{MaGra2}, \cite{MaGra3}.

In this paper we suggest to control the consistency of interaction
by exploiting the involutive form of the field equations. As we
explain in the next section, any field theory model can be
equivalently formulated in the involutive form, and the result of
such a reformulation is termed an \textit{involutive closure}.
Normally, the involutive closure retains all of the symmetry of the
original system that makes the method convenient for studying
covariant field equations. The involutive closure of the Lagrangian
equations is generally not Lagrangian anymore, even for such a
simple model as Proca equations. There is no pairing between gauge
identities and gauge symmetries\footnote{The gauge identities are
often termed as the Noether ones, because the second Noether theorem
states the isomorphism between the identities and symmetries. For
the general non-Lagrangian system of equations, including involutive
closure of Lagrangian one, there is no automatic Noether's
correspondence between symmetries and identities. That is why, we do
not use the term "Noether identity" to avoid the impression that it
is related to any gauge symmetry. } in the involutive closures of
the Lagrangian equations in general. In particular, even if the
Lagrangian does not have gauge symmetry, the involutive closure of
the Lagrangian equations can possess non-trivial gauge identities.
The gauge algebra of the involutive closure of the field equations,
with unrelated gauge symmetries and identities, turns out to be more
rich, in general, than the gauge algebra of the equivalent
non-involutive system. It is the structure of the gauge algebra of
the involutive closure of the field equations that controls the
number of physical degrees of freedom. That is why, when the
interactions are included, it is the stability of the gauge algebra
of the involutive closure of the field equations that ensures the
consistency of the theory.

The idea of involution is well developed in the theory of ODE's and
PDE's, and it is effectively applied to a broad range of the
problems in various fields, see \cite{Seiler}. It has been never
systematically studied, however, as a tool for controlling the
consistency of interactions - to the best of our knowledge. In the
next section we explain some of the basic notions related to the
involutive systems which are relevant in the context of consistent
interaction problem. Our consideration, being at the physical rigor
level, skips subtleties of the theory of involutive systems (for a
rigorous review see \cite{Seiler}), because we focus at the issues
of gauge algebra that are underdeveloped in this field at the
moment.

\section{Involutive form of field equations}
By the order of a field equation we understand the maximal order of
the derivatives of the fields in the equation. The maximal order of
the equations in the system is said to be the order of the system.

\vspace{0.3 cm}

\noindent {\textbf{Definition.}}{ \textit{A system of order $t$ is
said involutive if any differential consequence of the order less
than or equal to $t$ is already contained in the system.}}

\vspace{0.3 cm}

In the theory of involutive PDE's \cite{Seiler}, the definition above is normally complemented by more strong and intricate requirements which are inessential here, given the physical rigor level accepted and the problem addressed, so we adopt this simple definition.

If the system of field equations is supplemented by all the
differential consequences of the orders lower or equal to the order
of the system, it becomes involutive. The system brought to
involution in this way is said to be the \textit{involutive closure}
of the original system. Obviously, the original system is equivalent
to its involutive closure in the sense that all the solutions are
the same for both systems.

If the system of field equations can be brought to the involutive
closure just by inclusion derivatives of some of the equations, we
will also consider it involutive\footnote{In the theory of PDEs such
consequences are termed trivial integrability conditions. These are
inessential for the count of physical degrees of freedom.}, though
it is not, according to the
 definition above. This simplification is convenient, and it
does not lead to any contradiction as far as the problem of
consistent interactions is concerned.

To illustrate the distinctions between the involutive and
non-involutive form of field equations from the viewpoint of
variational principle and gauge algebra, consider the example.

\vspace{0.3cm} \noindent \textbf{{Example}:} \textit{Irreducible
massive spin 1 field in $d=4$ Minkowski space}. The field equations
of the model include the Klein-Gordon equations and the
transversality condition for the vector field,
\begin{eqnarray}
 \label{KG1} T_\mu & \equiv & (\Box -m^2) A_\mu = 0\, ,  \qquad ord(T_\mu) = 2\\
\label{Tr1} T_\bot & \equiv & \partial^\mu A_\mu=0 \, , \qquad
\qquad\quad
  ord\,(T_\bot)\,=\,1
\end{eqnarray}
The  order of equations is denoted by the symbol $ord$. This system
is involutive and it is obviously non-Lagrangian as it contains five
equations for the four component field $A_\mu$. There is no gauge
symmetry, though there exists a non-trivial gauge identity between
the equations (\ref{KG1}) and (\ref{Tr1}):
\begin{equation}\label{NI1}
    \partial^\mu T_\mu - (\Box -m^2)T_\bot \equiv 0 \, .
\end{equation}
It is useful to introduce the generator of gauge identity $L$, such
that the identity would read
\begin{equation}\label{Z1}
L^aT_a\equiv 0\, , \quad a=(\mu , \bot)\, , \qquad
L^\mu=\partial^\mu \, , \quad L^\bot=- (\Box -m^2)
\end{equation}
Define now a notion of the total order of gauge identity. Suppose
the $a$-th component of the generator of the gauge identity $L^a$ is
a differential operator of order $s_a$, and the order of the
equation $T_a$ is $t_a$. The order of the $a$-th identity component
$l_a$ is defined as the sum $l_a=s_a+t_a$. The total order of the
gauge identity generated by $L$ is defined as the maximum
\begin{equation}\label{ordZ}
ord (L)=\max_{a} \{l_a\} \, .
\end{equation}
In the case of the identity (\ref{NI1}), the differential operator
$L^\bot= -(\Box -m^2)$ has the order $s_\bot=2$, and it acts on the
first order equation $T_\bot$, so the order of the transverse
component of identity $l_\bot=2+1=3$. The order of the differential
operator $L^\mu=\partial^\mu$ is $s^\mu=1$, and it acts on the
second order equation $T_\mu$, so the order of this identity
component is the same, $l_\mu=1+2=3$. As we see, the total order of
the gauge identity (\ref{NI1}) is 3.

Alternatively, the same massive vector field model is described by the  Proca equations
\begin{equation}\label{Proca}
    P_\mu\equiv (\delta^\mu_\nu \, \Box -\partial_\mu\partial^\nu - m^2 \, \delta_\mu^\nu)A_\nu = 0 \, , \qquad \textrm{ord}(P_\mu)=2 \, ,
\end{equation}
that are Lagrangian. The Proca equations follow from (\ref{KG1}), (\ref{Tr1}):
$$
P_\mu\equiv T_\mu - \partial_\mu T_\bot \, ,
$$
and vice versa, the Klein-Gordon and transversality equations follow from  (\ref{Proca}),
 $$
 T_\bot\equiv - m^{-2}\partial^\mu P_\mu \, , \qquad T_\mu \equiv (\delta^\mu_\nu  - m^{-2} \partial_\mu\partial^\nu  )P_\nu \, .
$$
Notice that the Proca equations, being of the second order, have the
first order differential consequence - the transversality condition
(\ref{Tr1}). This means that the Proca system is not involutive.
Obviously, (\ref{Proca}) and (\ref{KG1}),(\ref{Tr1}) are equivalent
systems of equations. The involutive closure of Proca equations,
that includes (\ref{Proca}) and their first order consequence
(\ref{Tr1}), is an involutive non-Lagrangian system that contains
the gauge identity of the third order:
\begin{equation}\label{NIPro}
    \partial^\mu P_\mu  + m^2 T_\bot\equiv 0
\end{equation}
One can observe that the involutive form of the spin one equations,
being non-Lagrangian, has a gauge identity, so its gauge algebra is
non-trivial. Quite opposite, the Lagrangian Proca equations
equations are not involutive and have trivial gauge algebra, without
any gauge identity.

\vspace{0.2 cm}

A similar conclusion as in the above example holds in general: if
the Lagrangian equations are not involutive, their involutive
closure, being a non-Lagrangian system, will have the gauge algebra
with more independent gauge identity generators than the original
Lagrangian system has had. Notice that if the Lagrangian system has
a gauge symmetry, its involutive closure will obviously have the
same symmetry, so the gauge algebra of the involutive closure will
always have the original gauge symmetry as a subalgebra.

Running a couple steps ahead, notice that the structure of the gauge
algebra of involutive system unambiguously defines the number of
physical degrees of freedom $\mathcal{N}$ by the following formula:
\begin{equation}\label{DoF}
    \mathcal{N}=\sum_{k=0}^\infty k(t_k - l_k - r_k) \, .
\end{equation}
Here $t_k$ is a number of equations of order $k$ in the involutive
system, $l_k$ is the number of  gauge identities of $k$-th total
order, and $r_k$ is the number of gauge symmetry generators of $k$th
order\footnote{By the order of gauge symmetry generator we
understand the highest order of the derivative of the gauge
parameter involved in the gauge transformation of the fields. From
this point of view, in Lagrangian theory, where the gauge identities
and gauge symmetries have the same generators, the order of symmetry
can be different from the total order of the identity. This is
because the total order of identity is defined taking into account
the order of the equations it involves, while the order of the gauge
symmetry is indifferent to the order of equations.}. In this
formula, both the gauge symmetry and gauge identity generators are
supposed irreducible. The formula is derived in the Appendix, where
one can also find its extension to the case of reducible gauge
identities and symmetries. The number of physical degrees of freedom
is understood here as a number of the Cauchy data needed to define a
solution modulo gauge ambiguity. In variational case it coincides
with the dimension of the reduced phase space. So, if the
configuration space count is done, $\mathcal{N}$ has to be divided
by 2.

In the example above of the involutive equations for the massive
spin-1 in $d=4$, we have one first order equation and four second
order ones, so  $t_2=4, t_1=1$. There is one identity of the third
order $l_3=1$, and no gauge symmetry. Substituting these numbers
into the general formula (\ref{DoF}) one obtains $\mathcal{N}=1\cdot
1 +2\cdot 4 - 3\cdot 1=6$ that provides the correct answer, as the
massive vector field has 3 physical polarizations, and the physical
phase space is 6-dimensional.

Consider one more example illustrating relation (\ref{DoF}): a
scalar field $\phi(t,x)$ in two-dimensional space  subject to a pair
of the second order equations
\begin{equation}\label{phi}
  T_t\equiv\partial^2_t \phi (t,x) = 0 \, , \quad  T_x\equiv\partial^2_x \phi (t,x) = 0\, , \qquad
  ord (T_t)= ord (T_x) = 2 \, .
\end{equation}
There are no differential consequences of the second or lower
orders, so the system is in involution. There is the fourth order
gauge identity,
\begin{equation}\label{NIphi}
  \partial^2_t T_x-   \partial^2_xT_t \equiv 0 \, .
\end{equation}
With two second order equations ($t_2=2$) and one forth order
identity ($l_4=1$), relation (\ref{DoF}) brings zero as the number
of physical degrees of freedom. Let us directly check that it is a
correct count. The general solution reads
\begin{equation}\label{gsphi}
  \phi (t,x)  = A xt +Bx +Ct +D \, ,
\end{equation}
with $A,B,C,D$ being arbitrary integration constants. No arbitrary
function of $x$ or $t$ is involved in the general solution. This
means that the system, being $2d$ field theory, has no local
physical degrees of freedom.

Let us comment on  formula (\ref{DoF}). At first, we notice that the
relation is valid for involutive equations. So, prior to applying
this formula to a non-involutive system, one has to take an
involutive closure. The involutive closure can be taken in an
explicitly covariant way for covariant equations, it does not
require any $(3+1)$-splitting, unlike the Dirac-Bergman algorithm.
Also, it is important that the involutive closure can be found for
any system, be it Lagrangian or not, and it does not require any
special (e.g., first order) formulation. The orders of equations,
symmetries and identities can be easily found, so formula
(\ref{DoF}) provides a simple tool to covariantly control the number
of physical degrees of freedom. The second peculiarity of this
formula is that it is somewhat counterintuitive: it involves neither
the number of fields in the system, nor is it sensitive to the
number of zero order gauge symmetries, identities and equations.

 The problem of computing the number of the physical degrees of freedom has
 been thoroughly  studied in the theory of involutive systems \cite{Seiler},
 though the answer has been known in a completely different
 terms\footnote{The analysis is made by means of the theory of Cartan's differential systems,
 and the answers are formulated in terms of the Hilbert polynomials \cite{Seiler}.
 In principle, this way of counting physical degrees of freedom is sufficient,
 though it seems inconvenient in the context of relativistic field theories because it requires cumbersome
 and not always explicitly covariant computations.}, not explicitly appealing to the total orders of equations, identities and symmetries.
 In the context of the problem of consistent interactions, where the structure of the gauge algebra is the principal object to study,
 it is important to control the degrees of freedom in terms of the gauge algebra constituents.
 In the Appendix we deduce the formula (\ref{DoF}) proceeding from the definition accepted in the theory of involutive
 systems and based on the concept of strength of a system of equations. This concept was introduced by Einstein
 when he counted degrees of freedom in General Relativity \cite{E} and it has been further developed in many works
 (among which we mention \cite{Mar}, \cite{Sch},\cite{Mat},\cite{Su}) and
related to the count by Cauchy data \cite{Seiler}.

Notice that for Lagrangian equations, whenever they are
involutive from the outset (in this case, the gauge identity
generators coincide with the gauge symmetry ones), the receipt
(\ref{DoF}) for the degree of freedom count takes a special form
(\ref{DoFL2o1}), which has been well known before \cite{HTZ}. Notice
another special form of field equations where the asymmetry may
occur between gauge identities and symmetries: the unfolded
formulation of the higher spin fields (for review see \cite{V}).
This method utilizes the involutive form of the
unfolded equations, and it also benefits from the fact that all the
equations, symmetry and identity generators are of the first order.
The unfolded formalism involves, however, infinite number of field
equations, symmetries and identities. Formula (\ref{DoF}) can not
be immediately applied to the unfolded systems because all the
numbers in (\ref{DoF}) are supposed to be finite. In this case, the method
of $\sigma_{-}$-cohomology \cite{sigma1}, \cite{sigma2},
\cite{sigma3} provides a tool for the degree of freedom count.  The
$\sigma_{-}$-cohomology method allows one to pick out a
finite involutive subsystem such that the unfolded system will be its
infinite jet prolongation. The degree of
freedom count in the finite subsystem, being made by the formula
(\ref{DoF}), delivers the answer for the number of degrees of freedom
in the complete unfolded theory.

\vspace{0.2 cm}

Let us formulate now the key stages of the procedure we suggest for constructing consistent interactions, given the original free field equations:
\begin{enumerate}
\item The free system is to be brought to the involutive form.
\item All the gauge symmetries and identities are to be identified in the free involutive
system.
\item The interaction vertices are perturbatively included to comply with the three basic requirements in every order of coupling constant:
\begin{enumerate}
 \item The field equations have to remain involutive;
  \item The gauge algebra of the involutive system can be deformed, though the number of gauge symmetry and gauge identity generators remains
  the same as it has been in the free theory;
  \item The number of physical degrees of freedom, being established by  relation (\ref{DoF}), cannot change,
  though some of the involved orders can.
\end{enumerate}
\end{enumerate}
This procedure ensures finding all the consistent interaction
vertices, for any regular system of field equations.

\section{Gauge algebra of involutive systems}
As it has been already explained, if the Lagrangian system of field
equations is not involutive, its involutive closure will be
non-Lagrangian. It is the structure of the gauge algebra of the
involutive closure, not the original system of equations, that
controls number of physical degrees of freedom. That is why, one has
to study the gauge algebra of  dynamics in its non-Lagrangian
involutive form even if the Lagrangian exists. The exception is the
case where the Lagrangian equations are involutive from the outset.
In this special case, the known methods \cite{H1} work well, being
insufficient for general Lagrangians. Notice once again that the
non-involutive Lagrangian equations are not exceptional - many
models in physics are of this class, e.g. massive fields with spin.
This leads us to consider first the gauge algebra of the
non-Lagrangian dynamics.

The gauge algebra of the general (not necessarily Lagrangian) system
is known in the same details as in the Lagrangian case, and the
corresponding BRST complex is also well studied \cite{LS0},
\cite{KazLS} that allows one to systematically control all the
compatibility conditions. Below, we provide a simplified description
of the gauge algebra, without resorting to the corresponding
cohomological tools and leaving aside the higher compatibility
conditions, as these are less important in the context of
interaction problem.

\subsection{Algebra of gauge symmetries and identities in general field theory}

It is common to consider general structures of gauge algebra by
making use of the condensed notation, and we will follow this
tradition as it is convenient for presenting the general idea of the
method. In this notation, the fields are collectively denoted by
$\phi^i$, with $i$ being the condensed index that includes all the
discrete indices, and also the space-time points. For example, the
vector field $A^\mu (x)$ is indexed by $i=(\mu, x)$. Summation over
the condensed index implies integration over $x$.

In the condensed notation, any system of field equations reads
\begin{equation}\label{T}
    T_a(\phi)=0 \, , \qquad ord (T_a) = t_a,
\end{equation}
where $a$ is a condensed index, and $T_a$ is understood as a
function of the fields and their space-time derivatives up to some
finite order $t_a$. The discrete part of the condensed index $a$
labeling the equations is different, in general, from that of the
condensed index $i$ numbering the fields. For the Lagrangian
equations, $i$ coincides with $a$, though this is not true if the
involutive closure is considered instead of the original equations.
For example, the involutive closure of the Proca equations includes
both the original Lagrangian equations (\ref{Proca}) and the
transversality condition (\ref{Tr1}), so the indices belong to the
different sets: $i=(\mu, x)$, and $a=(\mu , \bot , x)$. For the
regular field equations, the order $t_a$ depends only on the
discrete part of the index $a$, not on the space-time point.

The general field equations can enjoy gauge symmetry transformations
\begin{equation}\label{R}
    \delta_\epsilon \phi^i= \epsilon^\alpha R_\alpha^i (\phi)\, ,  \qquad  \delta_\epsilon \left.T_a(\phi )\right|_{{T=0}} = 0 \, ,\quad \forall \epsilon^\alpha\, , \qquad ord (R_\alpha) = r_\alpha<\infty\,,
\end{equation}
where the gauge parameters $\epsilon^\alpha$ and generators
$R_\alpha^i (\phi)$ are understood in the sense of condensed
notation, i.e. the summation over $\alpha$  implies integration over
$x$. For example, in electrodynamics, $\delta_\epsilon A_\mu
(x)=\partial_\mu \epsilon (x)$, and hence $i=(\mu,x), \, \alpha=y$,
$R_\mu (x,y)= \partial_\mu \delta (x-y)$, so that $\delta_\epsilon
A_\mu (x)= \int dy R_\mu (x,y) \epsilon (y)$. Locality of the gauge
symmetry implies that the gauge generators $R_\alpha$ are the
differential operators of finite order, denoted $r_\alpha$, with
coefficients depending on the fields and their derivatives.

The condition (\ref{T}) defines the on-shell invariance of the equations that off-shell reads
\begin{equation}\label{RT}
    R_\alpha^i(\phi) \partial_i  T_a (\phi) = U_{\alpha a}^b (\phi ) T_b (\phi ) \, ,
\end{equation}
where the derivative $\partial_i$ is understood as variational and the structure coefficients $U_{\alpha a}^b (\phi )$ are supposed to be regular on shell.

The gauge identities can also take place for the general field
equations, being not necessarily related to the gauge symmetries
\begin{equation}\label{ZT}
    L_A^a (\phi )T_a (\phi) \equiv 0 \, , \qquad ord(L_A)= l_a \, .
\end{equation}
The gauge identity generators $L_A$ are supposed to be local
differential operators. The total order of the identity $ord (L_A)$
is defined by the order of the differential operator and the order
of the equation it acts on as explained in Section 3 below
relation (\ref{Z1}).

The gauge symmetry and gauge identity generators are considered
as trivial whenever they vanish on shell, that is
\begin{equation}\label{Triv}
   R_\alpha^i{}_{(triv)} = \rho_\alpha^{i \, a} (\phi) T_a \, ,  \qquad L_A^a{}_{(triv)}=\zeta_A^{ab} (\phi ) T_b\, , \quad \zeta_A^{ab} = - \zeta_A^{ba} \, ,
\end{equation}
where  $\rho$ and $\zeta$ can be arbitrary local differential operators of finite order
with the coefficients depending on the fields and their derivatives.

The sets $\{R_\alpha^i \}, \, \{L^a_A\}$ of the gauge symmetry and
gauge identity generators are supposed to be complete. The completeness means
that any other generator of gauge symmetry or identity, satisfying
(\ref{RT}) or (\ref{ZT}), must be a linear
combination of the generators from the given set modulo the trivial
ones (\ref{Triv}).

Let us discuss now the equivalence relations for the systems of
field equations. Two systems of field equations  are considered
as equivalent if they are related by a locally invertible change of
fields and/or by locally invertible linear combination of the left
hand sides of the equations. The admissible class of changes
of fields reads
\begin{equation}\label{Cphi}
\phi^i \rightarrow \phi^{\prime i}= \phi^{\prime i}(\phi, \partial \phi, \partial^2 \phi, \dots )\, ,
\end{equation}
where the existence is implied for the inverse change belonging to
the same class, i.e. the original fields $\phi$ have to be
unambiguously determined by (\ref{Cphi}) as functions of the fields
$\phi^\prime$ and their derivatives up to some finite order.

For example, consider the system of vector and scalar field. The change
$A^\prime_\mu= A_\mu -\partial_\mu\phi, \, \phi^\prime=\phi$ is
admissible as the  local inverse  change exists.

Given a set of fields, the admissible class of equivalence
transformations for the field equations reduces to the linear combining with invertible coefficients:
\begin{equation}\label{ET}
    T_a\sim T^\prime_a \,\, \Leftrightarrow \,\,  T^\prime_a =K^b_a(\phi) T_b \,\, ,\qquad  T_a =(K^{-1})^b_a(\phi) T^\prime_b \, ,
\end{equation}
where the elements of the transformation matrices $K$ and $K^{-1}$ are the differential operators of finite order.

Assuming the completeness of the generators of gauge symmetries and
gauge identities, one can derive the following consequences from the
relations (\ref{RT}), (\ref{ZT}):
\begin{equation}\label{RR}
R_\alpha^j\partial_j R_\beta^i - R_\beta^j\partial_j R_\alpha^i= U_{\alpha\beta}^\gamma R_\gamma^i + W_{\alpha\beta}^{i \, a}T_a \, ;
\end{equation}
\begin{equation}\label{RZ}
R_\alpha^j\partial_j  L_A^a = U_{\alpha A}^{ B}L_B^a +
W_{\alpha A}^{ab}T_b \, , \qquad W_{\alpha A}^{ab} = -
W_{\alpha A}^{ba} \, ,
\end{equation}
where $U,W$ are some structure functions. These relations have
further compatibility conditions involving higher structure
functions (see for details \cite{LS0}, \cite{KazLS}).  The existence of all
the higher structure functions and their locality have been proven in
 \cite{KLSloc}  under the condition that the generators $L, R$ and the structure function $U$ involved in (\ref{RT}) are all local.
 The corresponding existence theorem for Lagrangian theories has been known long before \cite{H2}.
 So, with the existence theorem, to ensure consistency of the field theory (\ref{T}), it is sufficient to fulfill relations (\ref{RT}), (\ref{ZT})
 with some differential operators $R, Z, U$ of finite order.

 \subsection{Gauge algebra and perturbative inclusion of interactions}
Consider involutive system of free field equations
\begin{equation}\label{T0}
T^{(0)}_a(\phi ) = 0 \, , \qquad ord (T^{(0)}_a)= t^{(0)}_a \, .
\end{equation}
As the free field equations are supposed to be linear, the
generators of gauge symmetries and gauge identities are the
differential operators with field-independent coefficients. With
this regard, relations (\ref{RT}), (\ref{ZT}) in the free theory
should have identically vanishing on-shell terms:
\begin{equation}\label{R0}
    R^{(0)i}_\alpha\partial_i T^{(0)}_a(\phi ) \equiv 0 \, , \qquad
    ord (R^{(0)}_\alpha )= r^{(0)}_\alpha \, ;
\end{equation}
\begin{equation}\label{R0}
L_A^{(0)a}T^{(0)}_a(\phi )\equiv 0 \, , \qquad
ord(L_A^{(0)})=l_A^{(0)}.
\end{equation}
Given the orders of the equations, gauge symmetries and identities,
the number of physical degrees of freedom in the free model,
${\mathcal{N}}^{(0)}$, is defined by (\ref{DoF}).

Perturbative inclusion of interaction is understood as a
deformation of the equations, identities and gauge symmetries by
nonlinear terms,
\begin{equation}\label{DeT}
    T^{(0)}_a \, \rightarrow \, T_a  = T^{(0)}_a +g T^{(1)}_a +g^2
    T^{(2)}_a + \dots \, ,
\end{equation}
\begin{equation}\label{DeR}
    R_\alpha^{(0)i}\, \rightarrow \, R_\alpha^{i} = R_\alpha^{(0)i} +g R_\alpha^{(1)i} +g^2
    R_\alpha^{(2)i}+ \dots \, ,
\end{equation}
\begin{equation}\label{DeZ}
    L_A^{(0)a}\, \rightarrow \, L_A^{a} = L_A^{(0)a} +g L_A^{(1)a} +g^2
    L_A^{(2)a}+ \dots \, .
\end{equation}
Here $g$ is a coupling constant considered as formal
deformation parameter, generators $L_A^{(1)a}$ and $R_\alpha^{(1)i}$ are linear
in fields and their derivatives; $T^{(1)}_a$, $L_A^{(2)a}$, and
$R_\alpha^{(2)i}$ are bi-linear, etc. Notice that in each order of
the deformation, the  orders of equations, identities and symmetries
can never decrease.

Now, we can give a more specific formulation of the consistency
conditions for the interactions than the general explanation in the
end of Section 3. The consistency of the interaction is provided if
the three conditions are fulfilled: (a) the system remains
involutive at each order in $g$; (b) the deformations do not break
the gauge algebra generated by relations (\ref{RT}), (\ref{ZT}),
though the structure functions can change in (\ref{RT}) as well as
the higher relations; (c) the  orders of the equations, symmetries,
and identities may increase, though the overall balance established
by relation (\ref{DoF}) cannot change, i.e., it is required
$\mathcal{N}{}^{(0)}=\mathcal{N}$ in every order in $g$. The
conditions (a) and (b) provide algebraic consistency of the system
with perturbatively included interactions, and (c) ensures that the
interacting system has the same number of physical degrees of
freedom.

Let us elaborate on the perturbative procedure of the interaction
inclusion. Suppose we have taken the most general ansatz for
$T^{(1)}_a, \, ord(T^{(0)}_a +g T^{(1)}_a)= t_a^{(1)}$, that does
not break involutivity, so (a) is fulfilled. Substituting this
ansatz into relations (\ref{RT}), (\ref{ZT}) and considering that in
the first order in $g$, we find the following relations between
$T^{(1)}, R^{(1)}, L^{(1)}$:
\begin{equation}
\label{RT1}  R_\alpha^{(0)i}
\partial_i T^{(1)}_a = U_{\alpha a}^{(1)b} T^{(0)}_b - R_\alpha^{(1)i}
\partial_i T^{(0)}_a \,,
\end{equation}
\begin{equation}\label{ZT1}
L^{(0)a}_AT^{(1)}_a + L^{(1)a}_AT^{(0)}_a =0.
\end{equation}
These relations impose nontrivial restrictions on the first order
interaction.

The first relation means that the free theory gauge transformation
has to leave the first order interaction on-shell invariant modulo
linear combination of variations of free equations. ``On-shell"
hereafter means on the free equations.

The second relation means that the free gauge identity generators
must leave the first order interaction on-shell invariant. Notice
that even if the model has no gauge symmetry, the involutive closure
of its equations can have non-trivial gauge identities. This means
that the conditions (\ref{ZT1}) are essential for consistency of
interactions even in the systems without any gauge invariance. If
the equations are Lagrangian and involutive from the
outset\footnote{As it has been already noticed in Section 3,
there are many non-involutive Lagrangian equations being of a
considerable interest in physics.}, relations (\ref{ZT1}) are
reduced to the the on-shell gauge invariance of the
cubic vertices in the Lagrangian. For the involutive Lagrangian
equations, the generators $R$ and $L$ coincide, and the relations
(\ref{RT1}) follow from (\ref{ZT1}) in this case. For general
system, including the involutive closure of the Lagrangian
equations, the relations (\ref{ZT1}) are not necessarily connected
with (\ref{RT1}). The relations (\ref{RT1}) are always first
examined in the Lagrangian case (see \cite{H1}-\cite{HGR}) to check
the first order consistency of the interaction in Lagrangian
dynamics. As is seen from  the explanations above, if the Lagrangian
field equations are not involutive, the first order consistency of
interaction  requires to independently impose the extra conditions
(\ref{ZT1}) on the vertices, because these are not necessarily
connected to the gauge symmetry of the Lagrangian.

Given a free model, the solution does not necessarily exist in any
theory for the first order consistency conditions (\ref{RT1}),
(\ref{ZT1}) imposed on the first order interactions $T^{(1)}$ and
the corresponding first order corrections to the gauge identity and
gauge symmetry generators $L^{(1)}, R^{(1)}$. If a solution
exists, it can be explicitly found as the system is linear.

The solutions for interactions are considered modulo ambiguities
related to the equivalence relations (\ref{ET}),
(\ref{Cphi}). In particular, a nonlinear change of fields in the
free equations is not considered as an interaction as well as a
linear combination of the free equations with field-dependent
coefficients.

If the order $t^{(1)}_a$ increases because of the first order
interactions, the orders of  gauge identity and symmetry generators
should also increase in a corresponding way to have the same number
of physical degrees of freedom (\ref{DoF}). If a
solution  to (\ref{RT1}), (\ref{ZT1}) exists with the correct
$\mathcal{N}$, one can proceed to the next order.

In the second order in $g$, the basic relations of the gauge algebra
(\ref{RT}), (\ref{ZT}) read
\begin{equation}\label{RT2}
    R_\alpha^{(0) i}\partial_i T^{(2)}_a + R_\alpha^{(1) i}\partial_i
    T^{(1)}_a + R_\alpha^{(2) i}\partial_i T^{(0)}_a =  U_{\alpha
    \,
    a}^{(1) b} T^{(1)}_b + U_{\alpha \, a}^{(2) b} T^{(0)}_b\,,
\end{equation}
\begin{equation}\label{ZT2}
    L_A^{(0) a} T_a^{(2)} + L_A^{(1) a} T_a^{(1)} + L_A^{(2) a}
    T_a^{(0)} = 0\,.
\end{equation}
In the first instance, these relations represent further
compatibility conditions for the first order interaction. Let us
explain that in the case of relation (\ref{ZT2}). On substituting
into (\ref{ZT2}) the expressions for $L_A^{(1) a}, T_a^{(1)}$
previously derived from (\ref{RT1}), (\ref{ZT1}), one has to get a
combination of the free theory gauge identity generators $L^{(0)}$
modulo free equations. This requirement is not automatically
fulfilled for any interaction derived from (\ref{RT1}), (\ref{ZT1}).
In some models it can be even possible that these relations are
inconsistent. In this case, one arrives at a no-go theorem for the
interaction. So, the second order relations (\ref{RT2}), (\ref{ZT2})
provide an additional selection mechanism for the first order
interactions. If this filter is passed by the first order
interactions, then relations (\ref{RT2}), (\ref{ZT2}) can be viewed
as a consistent algebraic system of linear equations defining the
second order contributions to the equations, gauge symmetry and
gauge identity generators: $T_a^{(2)}, R_\alpha^{(2)i}, L_A^{(2)
a}$. The solution for the second order interaction is to be
considered modulo the equivalence relations (\ref{Cphi}),
(\ref{ET}). In particular, the nonlinear changes of fields or
combinations of the lower order equations with field dependent
coefficients are not considered as interactions. If the solution for
$T_a^{(2)}$ involves the field derivatives of a higher order than
$T_a^{(1)}$ and $T_a^{(0)}$, then the  orders of the gauge
symmetries and gauge identities have to increase in the
corresponding way to provide the same number of physical degrees of
freedom according to relation (\ref{DoF}).

On substituting the second order interactions into the third order
expansion terms of the relations (\ref{RT}), (\ref{ZT}) one arrives
at the relations that represent the consistency conditions for
$T_a^{(2)}, R_\alpha^{(2)i}, L_A^{(2) a}$. This is much like the
relations (\ref{RT2}), (\ref{ZT2}) work for the previous order
equations and generators. Again, there can be either inconsistency
found at this stage, or one derives the third order interaction, and
the procedure repeats in the next order. Three different scenarios
are possible for further development of the iterative constructing
the interactions. The first is that the iterative analysis of the
expansion in $g$ of the conditions (\ref{RT}), (\ref{ZT}) will
terminate at certain order because of contradiction. This results in
a no-go theorem. The second is that starting from certain order all
the interactions become trivial. This results in a polynomial
interaction. The third option is that the procedure results in
nontrivial consistent interactions in every order. This leads to a
non-polynomial interaction.

In the case of involutive Lagrangian equations, this procedure
reduces to the commonly known method of inclusion interactions
between gauge fields (see \cite{H1} for a review). It has been
already mentioned that the involutive closure of non-involutive
Lagrangian equations is not Lagrangian anymore. Because of that, the
Lagrangian method does not apply to this case, whereas the method of
this section still works well, as well as for any other involutive
system of field equations. Our method exploits the same general idea
as the Lagrangian gauge approach: to include interactions by a
consistent deformation of the equations and the related gauge
algebra. The main distinctions are related to the fact that the
general gauge algebra of involutive system involves gauge identities
(\ref{ZT}) independently from gauge symmetries, and the involutive
form of equations allows one to effectively control the number of
physical degrees of freedom.

\section{Examples: Consistent self-interactions of massive fields of spin $1$ and $2$}

In this section we illustrate the general procedure of perturbative
inclusion of  consistent interactions described in Section 4 by
the examples of self-interactions for massive fields of spin one and
two. Applying this method we find all the consistent interaction
contributions (without higher derivatives) to the field equations of
the second order in fields. For the corresponding Lagrangians this
would correspond to the cubic vertices, though we find that some of
the admissible interactions do not follow from any Lagrangian.

\subsection{The massive spin 1 in d=4}
As it has been explained in previous section, there are equivalence
relations for the involutive field equations, so one can choose
various representatives from the equivalence class of the free
equations. For the spin 1, this choice is not unique either, as it
has been explained in Section 3. We choose the Proca equations
and the transversality condition as free involutive equations for
the spin 1,
\begin{equation}\label{ProTr}
T^{(0)}_\mu=\partial^{\nu}F_{\nu\mu}-m^2 A_\mu, \qquad
T^{(0)}_{\bot}=\partial^\nu A_\nu \, , \qquad ord(T^{(0)}_\mu)=2\, ,
\, \, ord( T^{(0)}_\bot ) =1\,.
\end{equation}
In this section, we adopt the following agreement for the strength
tensor and its dual: $ F_{\mu\nu}=\partial_\mu A_\nu-\partial_\nu
A_\mu,\quad
\widetilde{F}_{\mu\nu}=\frac{1}{2}\varepsilon_{\mu\nu\alpha\beta}F^{\alpha\beta}\,
. $ The choice of free equations in the form (\ref{ProTr}) is
slightly more convenient than the other equivalent options, e.g.
(\ref{KG1}), (\ref{Tr1}), because the second order equations in the
system (\ref{ProTr}) are Lagrangian that makes it easier to check
the consistency of the interactions.

Equations (\ref{ProTr}) admit the gauge identity whose
generator reads
\begin{equation}\label{ZPro}
L^{(0)\mu}=\partial^\mu,\qquad L^{(0)\bot}=m^2 \, , \qquad
L^{(0)\mu}T^{(0)}_\mu\, + \, L^{(0)\bot}T^{(0)}_\bot\equiv 0\, ,
\quad ord(L^{(0)})=3.
\end{equation}
So, the involutive form of the massive spin-$1$ free field equations
includes four second order equations and one of the first order together
with the third-order gauge identity between them. (The general
definition is provided by  relation (\ref{ordZ}) for the total order
of the gauge identity.)

The next step according to the general procedure of perturbative
inclusion of interactions, as described in Section 4, is to take the
most general covariant ansatz for the first order correction to the
field equations that does not break involutivity. Let us assume that
no higher order derivatives of the fields are included in the
interactions\footnote{This assumption does not restrict generality.
One can see that the inclusion of higher derivatives would
inevitably increase the number of physical degrees of freedom, as it
is defined by relation (\ref{DoF}). We omit the detailed proof of
this fact, though the simple evidence of that can be easily seen.
If, for example, the third order derivatives are included into
$T^{(1)}_\mu$, there will be four  equations of the third order, so
the positive contribution to $\mathcal{N}$ will increase by $4$.
There is only one gauge identity, so its total order should increase
at least by four to compensate that. To achieve such a growth of the
order of the identity, one has to raise the order of $T_\bot$. As
the order of the scalar equation raised, this will again increase
$\mathcal{N}$ with no way to compensate the latter growth of the
order. }. Then the most general ansatz reads

\begin{flalign}\label{T1Pro}
\begin{array}{l}
T^{(1)}_\mu=A^\alpha\Big(\rho_1F_{\mu\alpha}+\rho_3\widetilde{F}_{\mu\alpha}\Big)+
\rho_2\Big(A^\alpha\partial_\alpha A_\mu-T^{(0)}_\bot
A_\mu\Big)+\partial^\alpha A^\beta
 \Big(\rho_4\partial_\beta \widetilde{F}_{\alpha\mu}+\\[3mm]\qquad+\rho_5
\partial_\beta
F_{\alpha\mu}+\rho_{9}\partial_\alpha
\widetilde{F}_{\beta\mu}\Big)+\partial_{\mu}\Big(\rho_7\partial_\alpha
A_\beta
\partial^\alpha
A^\beta+\rho_8\partial_\beta A_\alpha
\partial^\alpha A^\beta+\rho_{6}
F_{\alpha\beta}\tilde{F}^{\alpha\beta}\Big)+\\[3mm]\qquad+\zeta_1 A_\mu
T^{(0)}_\bot+\zeta_2 \partial_\mu A^\alpha\partial_\alpha
T^{(0)}_\bot+\zeta_{3}
\partial^\alpha A_\mu\partial_\alpha T^{(0)}_\bot+
\zeta_4\widetilde{F}^{\alpha}_{\phantom{\alpha}\mu}\partial_{\alpha}T^{(0)}_\bot+
\zeta_5\partial_{\mu}\left(T^{(0)}_\bot\right)^2+ \\[3mm]\qquad+\zeta_{6}
T^{(0)}_\alpha\partial^\alpha A_\mu+\zeta_{7}
T^{(0)}_\alpha\partial_\mu
A^\alpha+\zeta_{8}\widetilde{F}^{\alpha}_{\phantom{\alpha}\mu}T^{(0)}_\alpha+\zeta_9
T^{(0)}_\bot T^{(0)}_\mu\,,\\[3mm]
T^{(1)}_\bot=\rho_{10}F^{\alpha\beta}F_{\alpha\beta}+\rho_{11}\partial^\beta
A^\alpha\partial_\alpha A_\beta+\rho_{12} F^{\alpha\beta}
\widetilde{F}_{\alpha\beta}+\rho_{13}m^2 A_\alpha
A^\alpha+\\[3mm]\qquad+(\zeta_{10}+\displaystyle{\frac{\rho_2}{m^2}})
\Big(T^{(0)}_\bot\Big)^2.
\end{array}&&\end{flalign}

The vertices with ten $\zeta$-coefficients are trivial as they are
reduced to a linear combination of the free equations with
field-dependent coefficients. Inclusion/exclusion of these vertices gives an example of the equivalence transformation
(\ref{ET}) with
\begin{flalign}\label{}\begin{array}{l}
\left(K_\zeta\right)_{\mu}^\nu=\delta^{\mu}_\nu+\zeta_6
\partial^\nu A_\mu+\zeta_7\partial_\mu A^\nu+
\zeta_8\widetilde{F}^{\nu}_{\phantom{\nu}\mu}+\zeta_9\delta^\nu_\mu
T^{(0)}_\bot, \qquad\left(K_\zeta\right)^{\bot}_\bot=1+\zeta_{10} T^{(0)}_\bot \,,\\[3mm]
\left(K_\zeta\right)_{\mu}^\bot=\zeta_1 A_\mu+\left(\zeta_2
\partial_\mu A^\alpha+\zeta_3\partial^\alpha A_\mu+
\zeta^{4}\widetilde{F}^{\alpha}_{\phantom{\alpha}\mu}\right)\partial_\alpha+
2\zeta_5\partial_\mu T^{(0)}_\bot,
\qquad\left(K_\zeta\right)_{\bot}^\nu=0\,.
\end{array}&&\end{flalign}
The inverse transformation has the form
$\left(K_\zeta\right)^{-1}=K_{-\zeta}+O(\zeta)$. We keep these terms
to simplify the control of trivial vertices in the next order of
interactions. Besides the trivial
terms, the most general covariant  quadratic ansatz (\ref{T1Pro})
includes a 13-parametrer set of the non-trivial contributions
with the coupling constants $\rho$.

Substituting the ansatz (\ref{T1Pro}) into the structure relations
(\ref{ZT1}), we obtain the following consistency conditions for
$T^{(1)}_\mu, T^{(1)}_\bot$:

\begin{flalign} \label{ZPro1}
\begin{array}{l}
L^{(0)\alpha}T^{(1)}_\alpha+L^{(0)\bot}T^{(1)}_{\bot}\equiv\partial^\alpha
T^{(1)}_\alpha+m^{2}T^{(1)}_\bot=
\\[3mm]\qquad=\Big(\displaystyle{\frac{\rho_1}{2}}+m^2(\rho_7+\rho_{10})\Big)F^{\alpha\beta}F_{\alpha\beta}+
\Big(\displaystyle{\frac{\rho_3}{2}}+m^2(2\rho_{6}+\rho_{12})\Big)
F^{\alpha\beta}
\widetilde{F}_{\alpha\beta}+\\[3mm]\qquad+\Big(\rho_2+m^2(2\rho_8+2\rho_7-\rho_5+\rho_{11})\Big)\partial^\beta
A^\alpha\partial_\alpha A_\beta
+m^2\Big(\rho_1+m^2\rho_{13}\Big)A_\alpha A^\alpha+\\[3mm]\qquad+\Big(-\displaystyle{\frac{\rho_{9}}{2}}+2\rho_{6}\Big)\partial_\gamma
F_{\alpha\beta}\partial^\gamma\tilde{F}^{\alpha\beta}+2\rho_7\partial_{\gamma}\partial_\beta
A_\alpha
\partial^{\gamma}\partial^\alpha A^\beta+
2\rho_8\partial_{\gamma}\partial_\alpha
A_\beta\partial^\gamma\partial^\alpha
A^\beta+\\[3mm]\qquad+\rho_{1}A^\alpha T^{(0)}_\alpha+\partial^{\alpha}A^{\beta}\Big(2\rho_7\partial_{\alpha}T^{(0)}_\beta+
(2\rho_8-\rho_5)\partial_{\beta}T^{(0)}_\alpha+2(\rho_7+\rho_8)\partial_{\alpha}\partial_\beta
T^{(0)}_\bot\Big)+\\[3mm]\qquad+4\rho_{6}\widetilde{F}^{\alpha\beta}\partial_\alpha T^{(0)}_\beta+O(\zeta)=0\qquad (\mathrm{mod}\phantom{a} T^{(0)})\,.
\end{array}
 && \end{flalign}
The $\zeta$-terms vanish  on-shell, and for this reason, the
parameters $\zeta$ remain arbitrary at this stage. The consistency
requirement (\ref{ZPro1}) imposes seven conditions on thirteen
interaction parameters $\rho$:
\begin{flalign}
\begin{array}{l} \rho_7=0,\qquad \rho_8=0,\qquad
\rho_{9}=4\rho_{6},\qquad\rho_{10}=-\displaystyle{\frac{\rho_{1}}{2m^2}},\qquad\rho_{11}=\rho_5-\displaystyle{\frac{\rho_2}{m^2}}\,,\\[3mm]
\rho_{12}=-\displaystyle{\frac{\rho_3}{2m^2}}-2\rho_{6},\qquad
\rho_{13}=-\displaystyle{\frac{\rho_{1}}{m^2}}\, .
\end{array}
 && \end{flalign}
Obviously, six parameters $\rho_1,\dots, \rho_6$ remain arbitrary,
while the seven others are fixed by these relations. Having the
consistency conditions (\ref{ZPro1}) fulfilled, we arrive at the
following six-parametrer set of vertices:

\begin{flalign}\label{VP1}
\begin{array}{l}
T^{(1)}_\mu=A^\alpha\Big(\rho_1F_{\mu\alpha}+\rho_3\widetilde{F}_{\mu\alpha}\Big)+
\rho_2\Big(A^\alpha\partial_\alpha A_\mu-T^{(0)}_\bot
A_\mu\Big)+\partial^\alpha A^\beta
 \Big(\rho_4\partial_\beta \widetilde{F}_{\alpha\mu}+\\[3mm]\qquad+\rho_5
\partial_\beta
F_{\alpha\mu}\Big)+\rho_{6}\Big(4\partial^\alpha
A^\beta\partial_\alpha
\widetilde{F}_{\beta\mu}+\partial_\mu (F^{\alpha\beta}\widetilde{F}_{\alpha\beta})\Big)\,,\\[3mm]
T^{(1)}_\bot=-\displaystyle{\frac{\rho_{1}}{m^2}}\Big(\displaystyle{\frac{1}{2}}F^{\alpha\beta}F_{\alpha\beta}+m^2
A_\alpha
A^\alpha\Big)+\Big(\rho_5-\displaystyle{\frac{\rho_2}{m^2}}\Big)\partial^\beta
A^\alpha\partial_\alpha
A_\beta-\\[3mm]\qquad-\Big(\displaystyle{\frac{\rho_{3}}{2m^2}}+2\rho_{6}\Big) F^{\alpha\beta}
\widetilde{F}_{\alpha\beta}+\displaystyle{\frac{\rho_2}{m^2}}\left(T^{(0)}\right)^2\qquad\qquad\qquad
(\mathrm{mod}\phantom{a} \zeta).
\end{array}&&\end{flalign}
The corresponding contributions to the gauge identity generators read
\begin{flalign} L^{(1)\nu}=-\rho_1A^\nu+\left(\rho_5\partial^\nu
A^\alpha-4\rho_{6}\widetilde{F}^{\alpha\nu}\right)\partial_\alpha,\qquad
L^{(1)\bot}=0\,.&& \end{flalign}
Consider now the problem of
compatibility of the first-order interactions at the next order.
Following the general prescription of Section 4, we have to
substitute the first-order gauge identity generators and equations
$L^{(1)}, T^{(1)}$ obtained above, into relations (\ref{ZT2}) and
examine their consistency. We have
\begin{flalign}\label{ZT2Pro}
\begin{array}{l}
L{}^{(1)\alpha}T{}^{(1)}_\alpha=-\rho_{1}\rho_{2}\Big(\displaystyle{\frac{1}{2}}A^\alpha\partial_\alpha
A^2-A^2 T^{(0)}_\bot\Big)-\rho_1A^{\nu}\partial^\alpha A^\beta
 \Big(\rho_4\partial_\beta \widetilde{F}_{\alpha\nu}+\rho_5
\partial_\beta
F_{\alpha\nu}+\\[3mm]\qquad +4\rho_{6}\partial_\alpha
\widetilde{F}_{\beta\nu}\Big)-\rho_1\rho_{6}A^\nu\partial_\nu
(F^{\alpha\beta}\widetilde{F}_{\alpha\beta})+\left(\rho_5\partial^\nu
A^\alpha-4\rho_{6}\widetilde{F}^{\alpha\nu}\right)\partial_\alpha
T^{(1)}_\nu
\end{array}&&\end{flalign}
As is seen from (\ref{ZT2}) the first order interactions $L^{(1)},
T^{(1)}$, having the form (\ref{ZT2Pro}) with six parameters
involved, will be compatible in the second order if there exist
functions $T^{(2)}_\alpha,\, T^{(2)}_\bot$ such that
$ord(T^{(2)}_\alpha)\leq 2$, $ord(T^{(2)}_\bot)\leq 1$ and the
following conditions are fulfilled:
\begin{equation}\label{ZT2Prog}
L^{(1)\alpha}T^{(1)}_\alpha+\partial^\alpha
T^{(2)}_\alpha+m^2T^{(2)}_\bot=0 \qquad\qquad\qquad (\text{mod
}T^{(0)}).
\end{equation}
On substituting (\ref{ZT2Pro}) into (\ref{ZT2Prog}), one can find that
no obstructions occur to the existence of $T^{(2)}_\alpha, \,
T^{(2)}_\bot$ with appropriate orders of field derivatives. For
example, we can always take
\begin{flalign}\label{}\begin{array}{l}
T^{(2)}_\mu=-\left(\rho_5\partial^\beta
A_\mu-4\rho_{6}\widetilde{F}_\mu^{\phantom{\mu}\beta}\right)T^{(1)}_\beta+
\rho_1 A^{\beta}\partial^\alpha A_\mu \Big(\rho_4
\widetilde{F}_{\alpha\beta}+\rho_5
F_{\alpha\beta}\Big)+\\[3mm]\qquad+\rho_1\rho_{6}\Big(4A^\beta\partial_\mu
A^{\alpha}\widetilde{F}_{\alpha\beta}+A_\mu
F^{\alpha\beta}\widetilde{F}_{\alpha\beta}\Big)\,,\\[3mm]
T^{(2)}_\bot=\displaystyle{\frac{\rho_{1}\rho_{2}}{m^2}}\Big(\displaystyle{\frac{1}{2}}A^\alpha\partial_\alpha
A^2-A^2
T^{(0)}_\bot\Big)-\displaystyle{\frac{\rho_1}{m^2}}\partial_\beta
A^\nu\partial^\alpha A^\beta\Big(\rho_4
\widetilde{F}_{\alpha\nu}+\rho_5 F_{\alpha\nu}\Big)\,.
\end{array}&&\end{flalign}
This means that the six-parametrer set of the first-order
interactions (\ref{VP1}), being the general solution to the first-order
condition (\ref{ZT1}),  admits a consistent extension to the
second order without any restriction on the parameters
$\rho$. We will not further elaborate here on the most general
interactions of the higher orders, although the method allows one to
study the issue in its full generality in any order, as it can be
seen from the first-order example. Instead, we will just notice some
special cases, where the perturbative procedure of interaction
inclusion can be consistently interrupted  already at the second
order level.

At first, notice that if the parameters $\rho$ are chosen in such a
way that $L^{(2)}_\bot=0$ and $T^{(2)}_\mu=0$, the identity will be
consistent without higher order contributions, i.e., with
$T^{(n)},L^{(n)}=0,n>2$. The corresponding first-order vertices are
called self-consistent.

Two special combinations of the parameters are possible that
define inequivalent self-consistent first-order interactions:
\begin{enumerate}
    \item $\rho_1\,,\rho_2\,,\rho_3\neq 0,\qquad\rho_4=\rho_5=\rho_{6}=0$ that corresponds to the equations at most
        cubic in fields. A further specialized option $\rho_1\rho_2=0$
        leads to the at most quadratic in fields interactions;
 \item $\rho_2\,,\rho_3\,,\rho_4\neq 0,\qquad\rho_1=\rho_5=\rho_{6}=0$ results in quadratic interaction.
\end{enumerate}
In the first case, the corresponding equations read
\begin{flalign}\label{S-CPro1}\begin{array}{l}
T_\mu=\partial^\alpha
F_{\alpha\mu}-m^2A_\mu+A^\alpha\Big(\rho_1F_{\mu\alpha}+\rho_3\widetilde{F}_{\mu\alpha}\Big)+
\rho_2\Big(A^\alpha\partial_\alpha A_\mu-T^{(0)}_\bot
A_\mu\Big),\\[3mm]
T_\bot=\partial^\alpha
A_\alpha-\displaystyle{\frac{\rho_{1}}{m^2}}\Big(\displaystyle{\frac{1}{2}}F^{\alpha\beta}F_{\alpha\beta}+m^2
A_\alpha
A^\alpha\Big)-\displaystyle{\frac{\rho_2}{m^2}}\Big(\partial^\beta
A^\alpha\partial_\alpha A_\beta-\left(T^{(0)}\right)^2\Big)
-\\[3mm]\qquad-\displaystyle{\frac{\rho_{3}}{2m^2}} F^{\alpha\beta}
\widetilde{F}_{\alpha\beta}+\displaystyle{\frac{\rho_{1}\rho_{2}}{m^2}}\Big(\displaystyle{\frac{1}{2}}A^\alpha\partial_\alpha
A^2-A^2 T^{(0)}_\bot\Big),\\[3mm]
\hspace{3cm}L^{\alpha}=\partial^\alpha-\rho_1A^\alpha,\qquad L^\bot=m^2.
\end{array}&&\end{flalign}
The second item results in a different self-consistent interaction
of the first order
\begin{flalign}\label{S-CPro2}\begin{array}{l}
T_\mu=\partial^\alpha F_{\alpha\mu}-m^2A_\mu+\rho_3
A^\alpha\widetilde{F}_{\mu\alpha}+
\rho_2\Big(A^\alpha\partial_\alpha A_\mu-T^{(0)}_\bot
A_\mu\Big)+\rho_4\partial^\alpha A^\beta\partial_\beta \widetilde{F}_{\alpha\mu},\\[3mm]
T_\bot=\partial^\alpha
A_\alpha-\displaystyle{\frac{\rho_2}{m^2}}\Big(\partial^\beta
A^\alpha\partial_\alpha A_\beta-\left(T^{(0)}\right)^2\Big)
-\displaystyle{\frac{\rho_{3}}{2m^2}} F^{\alpha\beta}
\widetilde{F}_{\alpha\beta},\\[3mm]
\hspace{4cm}L^{\alpha}=\partial^\alpha,\qquad L^\bot=m^2\,.
\end{array}&&\end{flalign}
These two different quadratic interactions, being self-consistent as
such, can be complemented by the higher order interactions. The more
general quadratic interactions (\ref{T1Pro}) need cubic
corrections to ensue consistency. Though such corrections exist, as
we have explained above, they can be inconsistent in the next
order of interaction.

Notice that the three-parameter sets of self-consistent
interactions (\ref{S-CPro1}), (\ref{S-CPro2}), being the most
general in this class, are not necessarily compatible with
variational principle, though the free theory admits Lagrangian
formulation.  One can see that only the one-parameter family of the
vertices (\ref{S-CPro1}), (\ref{S-CPro2}) is variational. It is the
case of $\rho_2=-\rho_1=g$ and the other $\rho$'s and $\zeta$'s
vanishing. The corresponding Lagrangian reads
$$\mathcal{L}=\mathcal{L}^{(0)}+\mathcal{L}^{(1)}=-\frac{1}{4}F^{\alpha\beta}F_{\alpha\beta}-\frac{m^2}{2}A^\alpha A_\alpha-
\frac{g}{2}\partial^\alpha A_\alpha A^\beta A_\beta\,.
$$
All other vertices of (\ref{S-CPro1}), (\ref{S-CPro2}) do not follow from variational
principle. This demonstrates that the general class of the
consistent interactions can be much broader than that of Lagrangian
ones. This fact can also mean that some of the no-go theorems for the
interactions, known in the Lagrangian framework, may be
bypassed if the requirement is relaxed for the vertices to be
variational.

\subsection{The massive spin 2 in d=4}
The irreducible spin-2 massive field theory can be described by
a traceless, symmetric, rank-2 tensor field $h_{\mu\nu}$
subject to the Klein-Gordon equations and the transversality condition
\begin{equation}\label{KG2}
   T^{(0)}_{\mu\nu}\equiv(\Box-m^2)h_{\mu\nu}=0,\quad
T^{(0)}_\nu\equiv\partial^\nu h_{\mu\nu}=0\,,\qquad
ord(T^{(0)}_{\mu\nu})=2,\quad ord(T^{(0)}_\mu)=1\,.
\end{equation}
These equations are involutive as there are no low order
differential consequences. Unlike spin 1, the equations are
inequivalent to any Lagrangian system formulated in terms of the
original irreducible field. The Fierz-Pauli Lagrangian \cite{FP}
that involves auxiliary scalar field and traceless tensor
$h_{\mu\nu}$ leads to the equations that are equivalent to
(\ref{KG2}). The Fierz-Pauli equations (FPE) are not involutive. Their
involutive closure is not Lagrangian and it has a more
complex structure than the equations (\ref{KG2}) formulated without
any auxiliary field. A similar picture is observed for all the
higher-spin massive fields. The Lagrangian formulation due to Singh and
Hagen \cite{SH} needs auxiliary fields, that makes
the system non-involutive. The involutive closure of the Sing-Hagen
equations is not Lagrangian anymore, and is more complex than the
system of Klein-Gordon equations and the transversality condition for the
traceless tensors. The aim of this subsection is to demonstrate by
the example of the spin-2 field that the minimal formulation of the
irreducible field equations, involving just the mass shell and
transversality conditions, is sufficient for iterative construction
of consistent interactions. Though this formulation is not
Lagrangian, it admits quantization and can enjoy all the other
advantages of Lagrangian formalism, including Noether's
correspondence between symmetries and conserved currents. The matter
is that the model admits a Lagrange anchor. As is known, the
Lagrange anchor \cite{KazLS}, being identified for not necessarily
Lagrangian field equations, allows one to path-integral
quantize the theory \cite{KazLS}, \cite{LS1},
\cite{LS2}, and also to connect symmetries with conservation laws
 \cite{KLSloc}, \cite{KLS1}.

Prior to seeking for consistent interactions, we have to
identify the gauge identity and gauge symmetry generators for the
free field equations (\ref{KG2}). The model has no gauge symmetry
and there exists four third-order gauge identities. The generators
are given by
\begin{equation}\label{Z0S2}
L^{(0)\mu\nu}_\alpha=\frac{1}{2}\left(\delta^\mu_\alpha\partial^\nu+\delta^\nu_\alpha\partial^\mu\right)\,,\qquad
L^{(0)}{}^\nu_\alpha=-(\Box-m^2)\delta^\mu_\alpha,
\end{equation}
\begin{equation}\label{Z0T0S2}
 L^{(0)\mu\nu}_\alpha T^{(0)}_{\mu\nu}+ L^{(0)}{}^\nu_\alpha T^{(0)}_{\nu}\equiv 0   \, , \qquad ord(L^{(0)})_\alpha=3\,.
\end{equation}
Following the general procedure of Section 4, to switch on the first
order interactions, one has to find the quadratic vertices
$T^{(1)}_{\mu\nu}, T^{(1)}_{\nu}$ such that the identities
(\ref{ZT1}) hold with $T^{(0)}_{\mu\nu}, T^{(0)}_{\mu},
L^{(0)\mu\nu}_\alpha  L^{(0)\mu}_\alpha$ defined by (\ref{KG2}),
(\ref{Z0S2}).

We do not study the most general case, restricting the quadratic
vertices by the ansatz with at most two derivatives in every
term:\footnote{ In the most general case, the terms can appear with
two second-order derivatives and four derivatives in total. The
interactions of this type are usually considered abnormal, and we
do not study them here.}
\begin{flalign}\label{T1S2}
\begin{array}{l}
T^{(1)}_{\mu\nu}=\rho_{8}\left(\partial_\mu
h_{\alpha\beta}\partial_\nu
h^{\alpha\beta}-\displaystyle{\frac{1}{4}}\eta_{\mu\nu}\left(\partial
h\right)^2\right)+\rho_3 \left(\partial_\alpha
h_{\beta\mu}\partial^\alpha
h^{\beta}_{\phantom{\beta}\nu}-\displaystyle{\frac{1}{4}}\eta_{\mu\nu}\left(\partial
h\right)^2\right)+\\[3mm]\qquad+\rho_{9}\left(\partial_\mu
h_{\alpha\beta}\partial^\alpha
h^{\beta}_{\phantom{\beta}\nu}+\partial_\nu
h_{\alpha\beta}\partial^\alpha
h^{\beta}_{\phantom{\beta}\mu}-\displaystyle{\frac{1}{2}}\eta_{\mu\nu}\left(\widetilde{\partial
h}\right)^2\right)+\rho_6
h^{\alpha\beta}\partial_{\alpha}\partial_{\beta} h_{\mu\nu}+\\[3mm]\qquad+
\rho_4\left(\partial_\alpha h_{\beta\mu}\partial^\beta
h^{\alpha}_{\phantom{\alpha}\nu}-\displaystyle{\frac{1}{4}}\eta_{\mu\nu}\left(\widetilde{\partial
h}\right)^2 \right)+
\rho_7\left(h^{\alpha\beta}\partial_\nu\partial_\mu
h_{\alpha\beta}-\displaystyle{\frac{1}{4}}\eta_{\mu\nu}h^{\alpha\beta}\Box
h_{\alpha\beta}\right)+\\[3mm]\qquad+m^2\rho_5\left(h_{\alpha\mu}h^{\alpha}_{\phantom{\alpha}\nu}-\displaystyle{\frac{1}{4}}\eta_{\mu\nu}h^2\right)+\rho_1
h^{\alpha\beta}\left(\partial_{\nu}\partial_\alpha
h_{\beta\mu}+\partial_{\mu}\partial_{\alpha}h_{\beta\nu}-\displaystyle{\frac{1}{2}}\eta_{\mu\nu}\partial_{\alpha}T^{(0)}_\beta\right)\,,\\[3mm]
T^{(1)}_\nu=\partial^{\mu}\left(\rho_2 h_{\alpha\mu}
h^{\alpha}_{\phantom{\alpha}\nu}+\rho_{10}\eta_{\mu\nu}h^2\right) \,,
\end{array}&&\end{flalign}
where we used the following abbreviations:
$$ h^2=h_{\alpha\beta}h^{\alpha\beta},\qquad \left(\partial h\right)^2=\partial_\nu
h_{\alpha\beta}\partial^\nu h^{\alpha\beta},\qquad
\left(\widetilde{\partial h}\right)^2=\partial_\nu
h_{\alpha\beta}\partial^\alpha h^{\nu\beta}.$$ Notice that the
trivial (on-shell vanishing) terms are omitted in (\ref{T1S2}). All
the vertices are identically traceless. Following the general procedure
of Section 4, we substitute the ansatz (\ref{T1S2}) into the
relations  (\ref{ZT1}) and examine consistency,
\begin{flalign}\label{ZT1S2}
\begin{array}{l}
L^{(0)\alpha\beta}_\nu
T^{(1)}_{\alpha\beta}+L^{(0)\alpha}_{\nu}T^{(1)}_\alpha\equiv\partial^{\nu}T^{(1)}_{\mu\nu}-(\Box-m^2)T_\nu=\\[3mm]\qquad
=\partial^{\mu}Q_{\mu\nu}+\rho_{8} \partial_\nu h^{\alpha\beta}
T^{(0)}_{\alpha\beta}+\rho_{9}\partial^\alpha
h^{\beta}_{\phantom{\beta}\nu}T^{(0)}_{\alpha\beta}+\rho_{9}
\partial_\nu h^{\alpha\beta}\partial_\alpha
T^{(0)}_\beta-\\[3mm]\qquad-(\rho_{9}+\rho_1)
\partial^\alpha h^{\beta}_{\phantom{\beta}\nu}\partial_\alpha
T^{(0)}_\beta+\rho_6 h^{\alpha\beta}\partial_\alpha\partial_\beta
T^{(0)}_\nu-\rho_6\partial^\beta
h^{\alpha}_{\phantom{\alpha}\nu}\partial_\alpha T^{(0)}_\beta+
\rho_7 h^{\alpha\beta}\partial_\nu T^{(0)}_{\alpha\beta})+\\[3mm]\qquad+\rho_1
h^{\alpha\beta}\partial_\nu\partial_\alpha T^{(0)}_\beta+\rho_1
h^{\alpha\beta}\partial_{\alpha}T^{(0)}_{\beta\nu}-(\rho_{9}+\rho_1)m^2
h^{\beta}_{\phantom{\beta}\nu}T^{(0)}_\beta-\\[3mm]\qquad-\partial^{\mu}\Big(\rho_1\eta_{\mu\nu}\displaystyle{\frac{1}{2}}h^{\alpha\beta}\partial_{\alpha}T^{(0)}_\beta+
\rho_2\left(h^{\alpha}_{\phantom{\alpha}\mu}T^{(0)}_{\alpha\nu}+h^{\alpha}_{\phantom{\alpha}\nu}T^{(0)}_{\alpha\mu}\right)+\left(2\rho_{10}+
\displaystyle{\frac{\rho_7}{4}}\right)\eta_{\mu\nu}
h^{\alpha\beta}T^{(0)}_{\alpha\beta}\Big)\,,
\end{array}&&\end{flalign}
where
\begin{flalign}\label{Q}
\begin{array}{l}
Q_{\mu\nu}=\partial^{\alpha}h^{\beta}_{\phantom{\beta}\mu}\partial_\alpha
h_{\beta\nu}\Big(\rho_{9}+\rho_3+\rho_1-2\rho_2\Big)+\partial^{\beta}h^{\alpha}_{\phantom{\alpha}\mu}\partial_\alpha
h_{\beta\nu}\Big(\rho_4+\rho_6\Big)+\\[3mm]\qquad+\displaystyle{\frac{1}{4}}\eta_{\mu\nu}\left(\partial
h\right)^2\Big(\rho_{8}-\rho_3+2\rho_7-8\rho_{10}\Big)+\displaystyle{\frac{1}{4}}\eta_{\mu\nu}\left(\widetilde{\partial
h}\right)^2\Big(2\rho_1-\rho_4\Big)+\\[3mm]\qquad+
m^2h^{\alpha}_{\phantom{\alpha}\mu}h_{\alpha\nu}\Big(\rho_{9}+\rho_5+\rho_1-\rho_2\Big)+
\displaystyle{\frac{1}{4}}m^2\eta_{\mu\nu}h^2\Big(2\rho_{8}-\rho_5+\rho_7-4\rho_{10}\Big)\,.
\end{array}&&\end{flalign}
At first order, the consistency requires the expression
(\ref{ZT1S2}) to vanish modulo $T^{(0)}$. Each term in (\ref{Q}) being
independent (modulo a conserved tensor), so we arrive at the system
of equations restricting the interaction parameters
\begin{flalign}\label{}
\begin{array}{l} \rho_{9}+\rho_3+\rho_1-2\rho_2=0,\qquad
\rho_{8}-\rho_3+2\rho_7-8\rho_{10}=0,\qquad
\rho_4+\rho_6=0\,,\\[3mm]\rho_{9}+\rho_5+\rho_1-\rho_2=0,\qquad 2\rho_{8}-\rho_5+\rho_7-4\rho_{10}=0,\qquad
2\rho_1-\rho_4=0\,.
\end{array}&&\end{flalign}
The solution to this system reads
\begin{flalign}\label{}
\begin{array}{l}
\rho_{8}=-\displaystyle{\frac{\gamma}{3}},\qquad
\rho_{9}=\gamma-\rho_1,\qquad
\rho_3=2\rho_2-\gamma,\qquad\rho_4=2\rho_1,\\[3mm]
\rho_5=\rho_2-\gamma\,,\qquad
\rho_6=-2\rho_1,\qquad\rho_7=4\rho_0-\displaystyle{\frac{\gamma}{3}},\qquad
\rho_{10}=\rho_0-\displaystyle{\frac{\rho_2}{4}}\,,
\end{array}&&\end{flalign}
where $\gamma,\rho_0,\rho_1,\rho_2$ are arbitrary constants. The term with the
coefficient $\rho_2$ can be considered as trivial, because it is
generated by the diffeomorphism in the space of fields:
$$ h_{\mu\nu}\mapsto
h_{\nu\nu}^{'}=h_{\mu\nu}+\rho_2\Big(h_{\alpha\mu}h^{\alpha}_{\phantom{\alpha}\nu}-\frac{1}{4}\eta_{\mu\nu}h^2\Big)\,.$$
Finally, we conclude that the set of non-trivial vertices for
the massive spin-$2$ equations may involve at most $3$ parameters. The
consistent quadratic vertices are given by
\begin{flalign}\label{T1S2C}
\begin{array}{l}
T^{(1)}_{\mu\nu}=\gamma\Big[-\displaystyle{\frac{1}{3}}\partial_\mu
h_{\alpha\beta}\partial_\nu h^{\alpha\beta}-\partial_\alpha
h_{\beta\mu}\partial^\alpha
h^{\beta}_{\phantom{\beta}\nu}+\displaystyle{\frac{1}{3}}\eta_{\mu\nu}\left(\partial
h\right)^2+\partial_\mu h_{\alpha\beta}\partial^\alpha
h^{\beta}_{\phantom{\beta}\nu}+\\[3mm]\qquad+\partial_\nu
h_{\alpha\beta}\partial^\alpha
h^{\beta}_{\phantom{\beta}\mu}-\displaystyle{\frac{1}{2}}\eta_{\mu\nu}\left(\widetilde{\partial
h}\right)^2-\displaystyle{\frac{1}{3}}h^{\alpha\beta}\partial_\nu\partial_\mu
h_{\alpha\beta}+\displaystyle{\frac{1}{12}}\eta_{\mu\nu}h^{\alpha\beta}\Box
h_{\alpha\beta}-\\[3mm]\qquad-m^2h_{\alpha\mu}h^{\alpha}_{\phantom{\alpha}\nu}+\displaystyle{\frac{m^2}{4}}\eta_{\mu\nu}h^2\Big]+\rho_1
\Big[h^{\alpha\beta}\left(\partial_{\nu}\partial_\alpha
h_{\beta\mu}+\partial_{\mu}\partial_{\alpha}h_{\beta\nu}-\displaystyle{\frac{1}{2}}\eta_{\mu\nu}\partial_{\alpha}T^{(0)}_\beta\right)-\\[3mm]\qquad-
\partial_\mu h_{\alpha\beta}\partial^\alpha h^{\beta}_{\phantom{\beta}\nu}-\partial_\nu
h_{\alpha\beta}\partial^\alpha
h^{\beta}_{\phantom{\beta}\mu}-2h^{\alpha\beta}\partial_\alpha\partial_\beta
h_{\mu\nu}+2\partial_\alpha h_{\beta\mu}\partial^\beta
h^{\alpha}_{\phantom{\alpha}\nu}\Big]+\\[3mm]\qquad+\rho_2\Big[2\partial_\alpha
h_{\beta\mu}\partial^\alpha
h^{\beta}_{\phantom{\beta}\nu}-\displaystyle{\frac{1}{2}}\eta_{\mu\nu}\left(\partial
h\right)^2+m^2h^{\alpha}_{\phantom{\alpha}\mu}h_{\alpha\nu}-\displaystyle{\frac{m^2}{4}}\eta_{\mu\nu}h^2\Big]+\\[3mm]\qquad+
\rho_0\Big[4 h^{\alpha\beta}\partial_\nu\partial_\mu
h_{\alpha\beta}-\eta_{\mu\nu}h^{\alpha\beta}\Box h_{\alpha\beta}
\Big]\,,\\[3mm]
T^{(1)}_\mu=\partial^{\mu}\left(\rho_2 h_{\alpha\mu}
h^{\alpha}_{\phantom{\alpha}\nu}+\Big(\rho_{0}-\displaystyle{\frac{\rho_2}{4}}\Big)\eta_{\mu\nu}h^2\right)\,.
\end{array}&&\end{flalign}
Notice that two parameters $\gamma$ and $\rho_1$ are
associated with the conserved currents that do not contribute to the
transversality condition $T^{(1)}_\mu$.

The first-order deformation of the Noether identity generators
is given by
\begin{flalign}\label{Z1S2}
\begin{array}{l}
L^{(1)\alpha\beta}_\nu=\partial^{\mu}\Big(\mathcal{L}_{\mu\nu}^{\alpha\beta}\cdot\Big)+\Big(\rho_1-\gamma\Big)\partial^{(\alpha}h^{\beta)}_{\phantom{\beta)}\nu}+
4\rho_0 \partial_\nu h^{\alpha\beta}+\rho_1\delta_{\nu}^{(\alpha}T^{(0)\beta)}\,,\\[3mm]
L^{(1)\alpha}_{\nu}=\partial^{\mu}\Big(\mathcal{L}_{\mu\nu}^{\alpha}\cdot\Big)+\widetilde{\mathcal{L}}_\nu^\alpha\,,
\end{array}&&\end{flalign}
where the round brackets mean symmetrization in corresponding
indices and the following notation is used:
\begin{flalign}\label{Z1S2+}
\begin{array}{l}
\mathcal{L}_{\mu\nu}^{\alpha\beta}=\Big(\displaystyle{\frac{\gamma}{4}-\frac{\rho_2}{2}}-\rho_0\Big)\eta_{\mu\nu}h^{\alpha\beta}
+\Big(\rho_2-\rho_1\Big)h^{(\alpha}_{\phantom{\alpha}\mu}\delta^{\beta)}_{\phantom{\beta}\nu}+\rho_2h^{(\alpha}_{\phantom{\alpha}\nu}\delta^{\beta)}_{\phantom{\beta}\mu}\,,\\[3mm]
\widetilde{\mathcal{L}}^\alpha_\nu=\gamma\Big(\partial_\nu
T^{(0)\alpha}-T^{(0)}{}^{\alpha}_{\phantom{\alpha}\nu}\Big)+\rho_{1}\Big(2\partial^\alpha
T^{(0)}_\nu+T^{(0)\alpha}\partial_\nu
-\partial_\nu T^{(0)}_{\alpha}-2\delta^{\alpha}_{\phantom{\alpha}\nu}T^{(0)}_\beta\partial^\beta\big)\,,\\[3mm]
\mathcal{L}_{\mu\nu}^\alpha=\rho_{1}\Big(\partial_\nu
h^{\alpha}_{\phantom{\alpha}\mu}+2\delta^{\alpha}_{\phantom{\alpha}\nu}h^{\beta}_{\phantom{\beta}\mu}\partial_{\beta}-
2\partial^\alpha h_{\mu\nu}-h^{\alpha}_{\phantom{\alpha}\mu}\partial_{\nu}+\displaystyle{\frac{1}{2}}\eta_{\mu\nu}h^{\beta\alpha}\partial_{\beta}\Big)+\\[3mm]\qquad+
\gamma\Big(\partial_\mu h^\alpha_{\phantom{\alpha}\nu}-\partial_\nu
h^\alpha_{\phantom{\alpha}\mu}\Big)\,.
\end{array}&&\end{flalign}

As it follows from the general requirement (\ref{ZT2}), the next
order consistency of the vertices (\ref{T1S2C}) is only possible
under the following condition:
\begin{equation}\label{RS2}
L_{\alpha}^{(1)\mu\nu}T^{(1)}_{\mu\nu}+L_\alpha^{(1)\mu}T^{(1)}_\mu=\partial^\mu
\mathcal{Q}_{\alpha\mu}-m^2T^{(2)}_{\alpha}\,\qquad
(\mathrm{mod}\phantom{a} T^{(0)})\,,
\end{equation} where
\begin{equation}\label{Rdef}
\mathcal{Q}_{\alpha\mu}=\partial_{\mu}T^{(2)}_{\alpha}-T^{(2)}_{\alpha\mu}\,.
\end{equation}
Relation (\ref{Q}) is the compatibility condition for the second-order
vertices. We will seek for the solutions to these equations that obey conditions
$$ord (\mathcal{Q}_{\mu\nu})=2,\qquad ord (T^{(2)}_\mu)=1\,.$$
These are obviously consistent with the number of physical degrees of
freedom. The admissible choice is
$$
\mathcal{Q}_{\nu\mu}=\mathcal{L}_{\mu\nu}^{\alpha\beta}T^{(1)}_{\alpha\beta}+\mathcal{L}_{\mu\nu}^\alpha
T^{(1)}_{\alpha}+\Big(\rho_1-\gamma\Big)
\Big[h^{\beta}_{\phantom{\beta}\nu}\Big(T^{(1)}_{\beta\mu}-\partial_{\mu}
T^{(1)}_\beta\Big)+\partial_{\mu}h^{\beta}_{\phantom{\beta}\nu}
T^{(1)}_\beta\Big]\,.
$$
To get $\mathcal{Q}$ we integrate by parts and take into account
the identity (\ref{ZT1S2}):
$$
h^{\beta}_{\phantom{\beta}\nu}\partial^{\alpha}T^{(1)}_{\alpha\beta}=
h^{\beta}_{\phantom{\beta}\nu}\Big(\Box-m^2\Big)T^{(1)}_{\beta}=
\partial^\mu\Big[h^{\beta}_{\phantom{\beta}\nu}\Big(T^{(1)}_{\beta\mu}-\partial_{\mu} T^{(1)}_\beta\Big)+
\partial_{\mu}h^{\beta}_{\phantom{\beta}\nu} T^{(1)}_\beta\Big]\qquad (\mathrm{mod\phantom{a}}T^{(0)})\,.
$$
The on-shell vanishing gauge identity generators
$\widetilde{\mathcal{L}}_{\nu}^{\alpha}$ and
$\delta_{\nu}^{(\alpha}T^{(0)\beta)}$ can not affect the relation
(\ref{RS2}), so the question left reads: is it possible to
represent the remaining term $h^{\alpha\beta}\partial_\nu
T^{(1)}_{\alpha\beta}$ in the form (\ref{RS2}) for some
$\mathcal{Q}$? If the answer is affirmative, one should try to express
$\mathcal{Q}$ in the form (\ref{Rdef}) with on-shell traceless and
symmetric tensor $T^{(2)}_{\alpha\beta}$. If the answer is negative,
an obstruction for the second order vertex appears. We have
\begin{flalign}\label{R0R1}
\begin{array}{l}
    4\rho_0\partial_\nu h^{\alpha\beta}T^{(1)}_{\alpha\beta}=\partial^{\mu}\Big[4\rho_0\Big(4\rho_0-\displaystyle{\frac{\gamma}{3}}\Big)
    \partial_\nu h^{\beta}_{\phantom{\beta}\mu}h^{\sigma\tau}\partial_{\beta}h_{\sigma\tau}+
    8\rho_0\rho_1\partial_\nu h^{\beta}_{\phantom{\beta}\mu} h^{\sigma\tau}\partial_{\tau}h_{\sigma\beta}\Big]-\,\\[3mm]
    \qquad-8\rho_0\rho_1\partial_\nu h^{\alpha\beta}h^{\sigma\tau}\partial_\sigma\partial_\tau h_{\alpha\beta}+\ldots\qquad\qquad\qquad (\mathrm{mod\phantom{a}}T^{(0)})\,,
\end{array}&&\end{flalign}
where the dots denote the terms of order $1$ that may be included
into $T^{(2)}_\nu$. There is a  term of order $2$ in the r.h.s. of
(\ref{R0R1})
$$
-8\rho_0\rho_1\partial_\nu
h^{\alpha\beta}h^{\sigma\tau}\partial_\sigma\partial_\tau
h_{\alpha\beta}
$$
that does not reduce on shell to a total divergence. It cannot also
be absorbed by deformation of the transversality condition
$T^{(2)}_\mu$ because of the order restriction. This means that the
vertices  are inconsistent in the class with $ord(T^{(2)}_\mu)=1$
unless $\rho_0\rho_1=0$. Therefore there can be at most two
2-parameter families  of consistent interactions in the considered
class. This seems matching well the fact that the massive gravity
admits a 2-parameter family of Lagrangians \cite{HR1} that are
consistent from the viewpoint of Hamiltonian constrained analysis
\cite{HR2}. The detailed comparison, however, with complete
nonlinear equations of massive gravity \cite{HR1} is not
straightforward as the vertices for the field equations (\ref{KG2})
are deduced for the traceless tensor $h_{\mu\nu}$ without any
auxiliary field involved. The massive gravity equations involve the
tracefull tensor. In the free limit the massive gravity reduces to
the FP equations. In the FP theory, the trace vanishes on shell,
 leading to the equations (\ref{KG2}) for traceless tensor. For the
nonlinear equations of massive gravity, the explicit on-shell
exclusion of the auxiliary field is unknown at the moment, though
one should expect it is still possible (if it was impossible, the
theory would have a different number of degrees of freedom). Unless
the auxiliary field remains involved in the equations, the
corresponding vertices can not be immediately compared with the ones
in the equations formulated without auxiliary fields from outset.
The goal of this section, however, is not to perturbatively
re-derive the massive gravity vertices, but to exemplify the
involutive technique of finding interactions without a direct use of
Lagrangian, making use of a minimal set of fields.

\section{Concluding remarks}
Let us briefly discuss the results. In this paper, we propose a new
method that allows one to examine the consistency of interactions
for the general field theory models, be they Lagrangian or
non-Lagrangian. The method also provides a technique for
perturbative identification of all the admissible interactions,
given a free field model. The method requires first to bring the
field equations to the involutive form. Notice that the involutive
closure is always non-variational for variational non-involutive
equations. As far as the field equations are brought to the
involutive form, the gauge algebra is to be identified for the
equations.  The consistency of the gauge algebra is examined by
tools similar to those based on the BV formalism for Lagrangian
systems \cite{H1}, \cite{BH1} with three major generalizations. The
first generalization is that the consistency of gauge algebra is
examined for the involutive closure of the system of field
equations, not for the action functional which might be even
non-existent. The second is that the gauge identity generators are
involved in the gauge algebra of the involutive closure
independently from the gauge symmetries. The identity generators
impose their own consistency conditions that are not identified by
previously known method \cite{H1} even in the Lagrangian case. The
third is that the gauge algebra of the involutive system provides a
convenient receipt (\ref{DoF}), (\ref{DF-A}) for counting physical
degrees of freedom. This formula is applied to the involutive
equations in a covariant form, and it uniformly covers all
conceivable instances. Let us also mention that the formula
(\ref{DF-A}) has been derived in the Appendix in a more general
setting than it is actually utilized in the paper, as it also
applies to the case of reducible gauge symmetries and gauge
identities. Let us finally notice, that the involutive closure of
field equations admits a BRST embedding along the lines of
 \cite{KazLS}. From the viewpoint of the corresponding local BRST complex
 \cite{KLSloc}, the formula for the degree of freedom count (\ref{DoF}),
 (\ref{DF-A}) can get a natural cohomological interpretation.
 This issue will be addressed elsewhere.

\subsection*{Acknowledgements.} We appreciate useful discussions with
I.V. Tyutin and E.D. Skvortsov on various issues related to the
topic of this paper. The work is partially supported by the grant No
14.B37.21.0911 from the Ministry of Science and Education of Russian
Federation. A.A.Sh. is grateful for the support from Dynasty
Foundation, and S.L.L. is partially supported by the RFBR grant
11-01-00830-a.

\section{Appendix. \\ Physical degree of freedom count}
In this Appendix, we explain the origin of formula (\ref{DoF}) for
counting of the  physical degrees of freedom.

The starting point for deriving the formula (\ref{DoF}) is the
notion of the strength of differential equations introduced by
Einstein \cite{E}. Roughly, the strength is a number that measures
the size of the solution space. The ``stronger'' is the system of
differential equations, the smaller is its solution space. It turns
out that the numerical value of strength can be immediately related
with the number of Cauchy data needed to define the general solution
modulo gauge freedom, i.e., with the number of physical degrees of
freedom. The original Einstein's argumentation was not
mathematically rigor, its justification and explanation within the
modern theory of formal integrability can be found in book
\cite{Seiler}. For earlier discussions  of the concept of strength
of equation as well as numerous applications to the analysis of
relativistic field equations we refer the reader to \cite{Mar},
\cite{Sch}, \cite{Mat}, \cite{Su}.

Let us first explain the Einstein's concept of the strength of field
equations. Consider a set of fields $\phi^i$ on $d$-dimensional
space-time with coordinates $x^\mu$. Assuming the fields to be
analytical functions, we can expand them in Taylor series about some
point $x_0$:
\begin{equation}\label{EXP-A}
    \phi^i(x)=\sum_{p=0}^\infty
    \frac1{p!}\varphi^i_{\mu_1\cdots\mu_p}(x-x_0)^{\mu_1}(x-x_0)^{\mu_2}\cdots
    (x-x_0)^{\mu_p}\,.
\end{equation}
Let $N_p$ denote the total number of terms of $p$th order in the
expansion above. (The explicit expression for $N_p$ is given below.)
As far as the fields $\phi^i$ obey a system of PDEs, not all the
Taylor coefficients $\varphi^i_{\mu_1\cdots\mu_p}$ can remain
arbitrary. Denote $N'_p$   the number of monomials of order $p$ that
are left free in the general solution of the field equations.
Obviously, $N'_p < N_p$. On the other hand, if the field equations
enjoy a gauge symmetry, not all the solutions are physically
relevant; some of the monomials in the general solution come from
Taylor series for the gauge parameters. Modding out by the gauge
freedom, one can define the number $N''_p$ of gauge inequivalent
monomials of order $p$ entering the general solution. Now, the
number of physical degrees of freedom \textit{per
point}\footnote{Accordingly, the number of physical
{\textit{polarizations}} of the field $\phi^i$ is the half of
$\mathcal{N}$.} is given by
\begin{equation}\label{N-A}
    \mathcal{N}=\lim_{p\rightarrow \infty} \frac{p}{d-1}\frac{{N}''_p}{N_p}\,.
\end{equation}
This formula, that dates dates back to Einstein,  defines the number
of physical degrees of freedom as the growth of the number of
``physical monomials'' compared to the unconstrained ones.

%In nonlinear theories with gauge symmetries and identities, it might
%be rather problematic to directly compute the limit (\ref{N-A}),
%though Einstein  had done that explicitly for electrodynamics and
%General Relativity \cite{E}.

%Below, we demonstrate that the limit (\ref{N-A}), that defines the
%number of physical degrees of freedom, is uniformly defined in its
%turn by a certain numbers characterizing the field equations. These
%numbers are the quantities and total orders of field equations,
%gauge symmetries, and identities. This uniform way of counting the
%degrees of freedom seems much more simple, and physically
%transparent, than the direct computation of the limits of Taylor
%coefficients in the expansion of the fields, equations, and gauge
%transformations for every specific model.

Consider  a system of PDEs
\begin{equation}\label{PDE-A}
 T_a(\phi^i, \partial_\mu\phi^i,\ldots,
 \partial_{\mu_1}\cdots\partial_{\mu_m}\phi^i)=0\,,\qquad a=1,\ldots,
 t,
\end{equation}
governing the dynamics of fields $\phi^i$, $i=1,2\ldots, f$. The
order of these equations equals to $m$. Substituting the expansion
(\ref{EXP-A}) into the field equations (\ref{PDE-A}) and evaluating
the result at $x=x_0$, we get the system of algebraic equations
\begin{equation}\label{T-A}
    T_a(\varphi^i, \varphi^i_\mu, \ldots, \varphi^i_{\mu_1\cdots\mu_m})=0\,.
\end{equation}
These equations follow from (\ref{PDE-A}) by simply replacing the
partial derivatives of fields with the corresponding Taylor
coefficients. In general, the solution space for these equations can
be a very complicated algebraic variety containing
strata of different dimensions. So, it might be problematic to
choose the independent coefficients and compute their totality. But
the task is considerably simplified if one considers the conditions on the
higher order coefficients. After all, the lower order monomials do
not matter  for evaluating the limit (\ref{N-A}). Differentiating
(\ref{PDE-A})
 $k$ times by $x$'s and setting $x=x_0$, we
 obtain a set of algebraic equations of the form
\begin{equation}\label{SL-A}
    \mathcal{J}{\,}^{\nu_1\cdots\nu_{m+k}}_{ai\mu_1\cdots\mu_k}\varphi^i_{\nu_1\cdots\nu_{m+k}}+
    \mathcal{I}_{a\mu_1\cdots \mu_k}=0\,,
\end{equation}
where the functions $\mathcal{J}$'s and $\mathcal{I}$'s depend on
$\varphi$'s of order less than $m+k$. Thus, for each given $k$, we
get a system of linear inhomogeneous equations for the coefficients
$\varphi^i_{\nu_1\cdots\nu_{m+k}}$. The matrix
${\mathcal{J}}=\mathcal{J}_k$, defining the system, is called the
\textit{symbol matrix} of order $k$; it has the following structure:
\begin{equation}\label{S-A}
\mathcal{J}{\,}^{\nu_1\cdots\nu_{m+k}}_{ai\mu_1\cdots\mu_k}=
\mathcal{J}{\,}_{ai}^{(\nu_1\cdots\nu_{m}}\delta^{\nu_{m+1}}_{\mu_1}\cdots\delta^{\nu_{m+k})}_{\mu_{k}}\,,\qquad
\mathcal{J}{\,}_{ai}^{\nu_1\cdots\nu_{m}}=\frac{\partial
T_{a}(\varphi)}{\partial \varphi^i_{\nu_1\cdots\nu_m}}\,.
\end{equation}
Here the round brackets mean symmetrization of the indices enclosed
and the functions $T_a(\varphi)$ are given by the left hand side of
equation (\ref{T-A}). As is seen, the symbol matrix $\mathcal{J}_k$
of order $k$ is expressed in a very specific way trough the symbol
matrix of order $0$. The latter may be called the symbol matrix of
the field equations (\ref{PDE-A}). For linear differential equations
the symbol matrix $\mathcal{J}_0$ is just the highest-order or
principal part of the system.

We thus see that whatever the original system of field equations may
be, there is an integer $m$ such that the space of monomials of
order $p>m$ is determined by a finite system of linear inhomogeneous
equations with coefficients depending on $\varphi$'s of order
$\leq p$. The echelon form of the algebraic equations (\ref{SL-A})
suggests to solve them one after another, so that at each step one
deals with a {\textit{finite}} system of linear inhomogeneous
equations. This makes possible applying the usual theorems of linear
algebra to evaluate the solution space.

First of all, the number of linearly independent solutions to
equations (\ref{SL-A}) crucially depends on the rank of the symbol
matrix $\mathcal{J}_l$. The symbol matrix in its turn is the
function of the Taylor coefficients
$\{\varphi^i_{\mu_1\cdots\mu_j}\}_{j=0}^{m}$ constrained by the
algebraic equations (\ref{T-A}), so that the rank of $\mathcal{J}_k$
can suddenly change. To avoid this complication we will restrict
ourselves to those solutions of (\ref{T-A}) for which the rank of
the symbol matrix $\mathcal{J}_k(\varphi)$ is maximal. This means
that we consider only the general (opposite to singular) solutions
to the field equations. Following \cite{Seiler}, we will call
$\{\varphi^i_{\mu_1\cdots\mu_j}\}_{j=0}^{m}$ the \textit{principal
coefficients}, referring to the other Taylor coefficients as
\textit{parametric}.

By Kronecker-Capelli's criterion, the system of linear inhomogeneous
equations (\ref{SL-A}) is compatible iff each left null-vector
$\mathcal{K}$ of the symbol matrix $\mathcal{J}_k$ annihilates also
the inhomogeneous term $\mathcal{I}_k$. Clearly, the null-vectors of
the symbol matrix, if any, can always be chosen to be functions of
the principal coefficients alone. A crucial point is that the
compatibility criterion is automatically satisfied for differential
equations in involution. The reason is very simple: vanishing of the
function
$\mathcal{K}^{a\mu_1\cdots\mu_k}\mathcal{I}_{a\mu_1\cdots\mu_k}$
would otherwise give a nontrivial constraint on the lower order
coefficients that, in turn,  would be manifestation of a hidden
integrability condition. On the other hand, if the vector
$\mathcal{K}$ annihilates both the symbol matrix $\mathcal{J}_k$ and
the inhomogeneous term $\mathcal{I}_k$, then it defines an  identity
for the linear equations (\ref{SL-A}) and this identity  must
follow from a gauge identity for the original differential equations
(\ref{PDE-A}).

The general solution to an inhomogeneous linear system is given by
its partial solution plus the general solution to the corresponding
homogeneous system. In our case, the latter is completely determined
by the symbol matrix. Thus, to evaluate the size of the solution
space, we focus on the solutions to the homogeneous system. These
form a linear space, whose dimension is given by the number of
unknowns minus the rank of the symbol matrix. The number of unknowns
$\{\varphi^i_{\nu_1\cdots\nu_{m+k}}\}$ in (\ref{SL-A}) coincide with
the number $N_p$ of lineally independent monomials of order $p=m+k$.
It is easy to find that
\begin{equation}\label{NP-A}
    N_p=f\binom{p+d-1}{p}=f\frac{(p+d-1)! }{p!(d-1)!}\,.
\end{equation}
The rank of the symbol matrix $\mathcal{J}_k$ can be computed as the
difference between the number of equations (\ref{SL-A}) and the
number of left null-vectors of the matrix $\mathcal{J}_k$. The
former is expressed through the binomial coefficients as
\begin{equation}\label{NM-A}
   t \binom{k+d-1}{l}=  t \binom{p-m+d-1}{p-m}\,.
\end{equation}
As was explained above all left null-vectors for the symbol matrix
of involutive equations come from gauge identities. Each gauge
identity
\begin{equation}\label{LT}
\hat{L}^{a}T_a\equiv 0
\end{equation}
is defined by differential operators
$$
\hat{L}{}^a=\sum_{n=0}^{q'} \mathcal{L}^{a
\nu_1\cdots\nu_n}\partial_{\nu_1}\cdots
\partial_{\nu_n}\,,
$$
with coefficients depending on fields and their derivative up to
some finite order $j$. If the highest coefficients
$\{{\mathcal{L}}^{a\mu_1\cdots\mu_q}\}$ are not all equal to zero
identically, then the number $q'$ is called the \textit{order of the
gauge identity} (\ref{ZT}). Differentiating (\ref{LT}) $s$ times by
$x$'s  and setting $x=x_0$, we find
\begin{equation}\label{ZS-A}
\mathcal{L}^{a\mu_1\cdots\mu_{q'}}\mathcal{J}_{ai\mu_1\cdots
\mu_{s+{q'}}}^{\nu_1\cdots \nu_{m+s+q'}}\varphi^i_{\nu_1\cdots
\nu_{m+s+q'}}+ \cdots \equiv 0\,,
\end{equation}
where dots stand for the terms involving $\varphi$'s of order less
than $ m+s+q'$. If $s$ is large enough such that $m+s+q'>j$, then we
are lead to conclude that the coefficients at
$\varphi^i_{\nu_1\cdots \nu_{m+s+q'}}$ in (\ref{ZS-A}) must be zero
in order for the identity to hold. This implies that the symbol
matrix $\mathcal{J}_{s+q'}$ admits the set of null-vectors
$\{\mathcal{L}_{\lambda_1\cdots\lambda_s}\}$ of the form
$$
\mathcal{L}_{\lambda_1\cdots\lambda_s}^{a\mu_1\cdots\mu_{s+q'}}=\mathcal{L}^{a(\mu_1\cdots\mu_{q'}}
\delta^{\mu_{q'+1}}_{\lambda_1}\cdots\delta^{\mu_{p+s})}_{\lambda_s}\,,\qquad
\mathcal{L}_{\lambda_1\cdots\lambda_s}^{a\mu_1\cdots\mu_{s+q'}}\mathcal{J}_{ai\mu_1\cdots
\mu_{s+q'}}^{\nu_1\cdots \nu_{m+s+q'}}=0\,.
$$
All these null-vectors are linearly independent and their number is
given by
\begin{equation}\label{}
    \binom{s+d-1}{s}\,.
\end{equation}
In other words, each gauge identity of order $q'$  for the
involutive equations of order $m$ gives
\begin{equation}\label{}
    \binom{l-q'+d-1}{l-q}= \binom{p-q+d-1}{p-q}
\end{equation}
left null-vectors for the corresponding symbol matrix
$\mathcal{J}_k$ provided that $k$ is large enough. Here $p=k+m$ and
the number $q=q'+m$ is called the \textit{total order} of gauge
identity.

Now, suppose that the system (\ref{PDE-A}) involves equations of
different orders: $t_0$ equations of order zero (algebraic
equations), $t_1$ equations of the first order and so on. Let us
also assume that the system contains no hidden integrability
conditions and becomes involutive upon adjoining trivial
differential consequences. Then according to  (\ref{NM-A}) all these
differential equations give rise to
\begin{equation}\label{MQ-A}
 \sum_{n=0}^\infty t_n \binom{p-n+d-1}{p-n}
\end{equation}
linear equations for the parametric coefficients
$\varphi^i_{\mu_1\cdots\mu_p}$ with large $p$. Of course, only a
finitely many terms are different from zero in the above sum. Let us
further suppose that the field equations enjoy  $l_n$ gauge
identities of \textit{total} orders $n=0,1,2,\ldots$. If  all these
identities are independent (irreducible), then the linear equations
(\ref{SL-A}) for the parametric coefficients of order $p$ are
possessed of exactly
\begin{equation}\label{NQ-A}
   \sum_{n=0}^\infty l_n\binom{p-n+d-1}{p-n}
\end{equation}
dependencies  (left null-vectors) provided that  $p$ is large
enough. The difference between (\ref{MQ-A}) and (\ref{NQ-A}) is the
number of independent equations for unknowns
$\varphi^i_{\mu_1\cdots\mu_{p}}$. Subtracting this difference from
(\ref{NP-A}), we get the dimension of the solution space, that is,
the number of independent monomials of order $p$:
\begin{equation}\label{N'-A}
    N'_p=f\binom{p+d-1}{p}-\sum_{n=0}^\infty (t_n-l_n)\binom{p-n+d-1}{p-n}\,.
\end{equation}

Now, we should take into account gauge freedom. Each gauge
transformation has the form
\begin{equation}\label{R}
\delta_\epsilon \phi^i =\sum_{n=0}^{q}
\mathcal{R}^{i\mu_1\cdots\mu_{n}}\partial_{\mu_1}\cdots\partial_{\mu_n}\epsilon\,.
\end{equation}
In this expression, the coefficients $\mathcal{R}$'s  are functions
of the fields and their derivatives up to some finite order and the
infinitesimal gauge parameter $\epsilon$ is an arbitrary function of
$x$'s. The number $q$ is the order of the gauge transformation.  The
gauge invariance of the field equations (\ref{PDE-A}) implies that
\begin{equation}\label{TU-A}
\delta_\epsilon T_a = \hat{U}{}^b_a T_b
\end{equation}
for some matrix differential operator $\hat{U}$. Let us expand the
gauge parameter in Taylor series $$
 \epsilon(x)=\sum_{n=0}^\infty \frac1{n!}\varepsilon_{\mu_1\cdots\mu_n}(x-x_0)^{\mu_1}\cdots(x-x_0)^{\mu_n}\,.
 $$
Then $s$-fold differentiation of the equality (\ref{TU-A}) at  $x_0$
yields the identity
\begin{equation}\label{JR-A}
  \mathcal{J}_{ai}^{\mu_1\cdots\mu_m} \mathcal{R}^{i\mu_{m+1}\cdots \mu_{m+q}} \varepsilon_{\mu_1\cdots\mu_{m+q+s}}+ \cdots = 0\,.
\end{equation}
Here the dots stand for $\varepsilon$'s  of order less than $m+q+s$
and all $\varphi$'s are assumed to define a solution to the field
equations. Since all $\varepsilon$'s are arbitrary, we conclude that
the leading term in (\ref{JR-A}) must vanish separately. This
results in the  set $\{\mathcal{R}^{\nu_1\cdots \nu_{q+m+s}}\}$ of
right null-vectors for the symbol matrix $\mathcal{J}_s$:
$$
\mathcal{R}^{i\nu_1\cdots
\nu_{m+s+q}}_{\mu_1\cdots\mu_{m+l}}=\mathcal{R}^{i(\nu_1\cdots
\nu_{q}}\delta^{\nu_{q+1}}_{\mu_1}\cdots\delta^{\nu_{q+m+s})}_{\mu_{m+s}}\,,\qquad
{\mathcal{J}}_{ai\lambda_1\cdots{\lambda_{l}}}^{\mu_1\cdots\mu_{m+s}}\mathcal{R}^{i\nu_1\cdots\nu_{m+s+q}}_{\mu_1\cdots\mu_{m+s}}=0\,.
$$
These null-vectors span the space of dimension
$$
    \binom{m+s+q+d-1}{m+s+q}=\binom{p+q+d-1}{p+q}\,.
$$
In general, the system (\ref{PDE-A}) may enjoy several gauge
symmetry transformations. We let  $r_n$ denote the number of the
gauge transformations of order $n$. If all these gauge symmetries
are independent (irreducible), then they make
$$
\sum_{n=0}^\infty r_n\binom{p+n+d-1}{q+n}\,.
$$
coefficients of $\{\varphi_{\mu_1\cdots\mu_q}\}$ unphysical.
Subtracting this number from (\ref{N'-A}),  we get the number of
``physically distinguishable'' parametric coefficients of $p$th
order,
\begin{equation}\label{}
  N''_p=f \binom{p+d-1}{q}-\sum_{n=0}^\infty \left\{ (t_n-l_n)\binom{p-n+d-1}{p-n}+r_n\binom{p+n+d-1}{p+n}\right\}\,.
\end{equation}

Having computed  $N_p$ and $N''_p$ we are ready to evaluating the
limit (\ref{N-A}). Making use of the asymptotic expansion for the
binomial coefficients \cite{Mar}, \cite{Su}
$$
\binom{p\pm n+d-1}{p\pm n}=\binom{p+d-1}{p}\left\{1\pm
\frac{n}{p}(d-1)+ O\left(\frac{1}{p^2}\right)\right\}\,, \qquad
p\rightarrow \infty\,,
$$
we find
\begin{equation}\label{NN-A}
N''_p/N_p=(f-t+l-r) + \frac{(d-1)}{p}\sum_{n=0}^\infty
n(t_n-l_n-r_n)+O\left(\frac{1}{p^2}\right)\,.
\end{equation}
The numbers
$$
t=\sum_{n=0}^\infty t_n\,,\qquad l=\sum_{n=0}^\infty l_n\,,\qquad
r=\sum_{n=0}^\infty r_n
$$
coincide, respectively,  with the total number of
equations, gauge identities, and gauge symmetries. The leading term
of the expansion
\begin{equation}\label{D-A}
\Delta=f-t+l-r
\end{equation}
is called the \textit{compatibility coefficient}. Let us assume that
the system (\ref{PDE-A}) is \textit{absolutely compatible}, that
means  $\Delta=0$. Then comparing (\ref{NN-A}) with (\ref{N-A}), we
finally arrive at the desired formula for the physical degrees of
freedom
\begin{equation}\label{}
  \mathcal{N}=\sum_{n=0}^\infty n(t_n-l_n-r_n)\,.
\end{equation}

Vanishing of the compatibility coefficient $\Delta$ can be easily
established under the assumption that each right null-vector of the
symbol matrix $\mathcal{J}=\mathcal{J}_0$ originates from some gauge
symmetry. To do this,  we introduce the $n\times t$-matrix
$${J}_{ai}(p)=\mathcal{J}_{ai}^{\mu_1\cdots\mu_{m}}p_{\mu_1}\cdots p_{\mu_{m}}\,,$$ whose entries are polynomials in formal variables $p_\mu$, $\mu=1,\ldots, d$.  It then follows from (\ref{ZS-A}) that each gauge identity provides the left null-vector
$$
{L}^a(p) = \mathcal{L}^{a\mu_1\cdots \mu_{q'}}p_{\mu_1}\cdots
p_{\mu_{q'}}
$$
for the polynomial matrix ${J}(p)$, so that $L^a(p)J_{ai}(p)=0$.
Similarly, each gauge symmetry transformation (\ref{JR-A}) gives
rise to the polynomial vector
$$
R^i(p)=\mathcal{R}^{i\mu_1\cdots\mu_q}p_{\mu_1}\cdots p_{\mu_q}
$$
annihilating the matrix $J(p)$ on the right, that is,
$J_{ai}(p)R^i(p)=0$. The vanishing condition for the compatibility
coefficient (\ref{D-A}) can be written as
$$
t-l=f-r\,.
$$
It means that the rank of the rectangular matrix $J(p)$, being
computed by the number of left null-vectors, coincides with its rank
defined in terms of right null-vectors. Clearly, this equality takes
place for any matrix over an algebraic field, say $\mathbb{R}$ or
$\mathbb{C}$. It turns out that the same statement holds true for
the matrices over the ring of polynomials in $p$'s provided that the
null-vectors $L$'s and $R$'s are linearly independent (over the ring of polynomials in $p$'s) and span the
right and left kernel spaces of the matrix  $J$ (see, e.g.
\cite{Eis}).

There is also another, more direct, interpretation of the absolute
compatibility condition. It can be shown \cite{Seiler} that the
value $f-t+l$ defines the number of arbitrary functions of $d$
variables entering the general solution to the field equations
(\ref{PDE-A}). The equality $\Delta=f-t+l-r=0$ then implies that
\textit{all} these functional parameters owe their existence to the
gauge symmetries. To the best of our knowledge,  example of field
equations  has been yet unknown that would not be absolutely
compatible. Moreover, the results of \cite{LS} suggest that any
system of ODEs is absolutely compatible and the same is true for
two-dimensional field theories \cite{LS-PR}.   So, it is a very
plausible hypotheses that any reasonable  field theory is absolutely
compatible.

The above consideration can be extended to the field equations with
reducible gauge symmetries and/or identities. Without further ado we
just present the final formula for the physical degrees of freedom,
which might be deduced by appropriate adjustment of the derivation
in the irreducible case:
 \begin{equation}\label{DF-A}
    \mathcal{N}=\sum_{m,n=0} n\big(t_n - (-1)^m(l_n^m + r_n^m)\big) \, .
\end{equation}
Here $t_n$ is the number of the equations of order $n$; $l_n^m$ is
the number of gauge identities of the total order $n$, and the
reducibility order $m$; and $r_n^m$ is the number of gauge symmetry
transformations of the total order $n$ and reducibility order $m$.
The total order of the generator of gauge symmetry/identity is
defined inductively to be the sum of its order as a differential
operator and the total order of a generator it annihilates. It is
also assumed that the total order of the original gauge symmetry
generators coincides with their order as differential operators,
whereas the total order of gauge identities for the field equations
is given by the order of the corresponding generators plus the order
of equations they act on. For the field theories with irreducible
gauge symmetries of the first order, equality (\ref{DF-A}) was first
derived in \cite{Su}.

It is curious to note that the final formula (\ref{DF-A}), being
independent on  $d$, holds true for the one-dimensional systems as
well, whereas the original definition (\ref{N-A}) becomes
meaningless. The proof of (\ref{DF-A}) for $d=1$ requires a
different method, which is beyond the scope of this paper.

Let us concretize the formula (\ref{DF-A}) for
the special case of involutive Lagrangian second order equations as
this case has a common interest in field theory. For the Lagrangian
equations, the gauge symmetries and identities are generated by the
same operators. The total order of the identities, however, is
shifted by the order of the equations involved in, so the careful
adjustment of the general relation (\ref{DF-A}) for the second order
involutive Lagrangian equations leads to the following count of
physical degrees of freedom:
\begin{equation}\label{DoFL2}
    \mathcal{N}= 2 \left( t_2 + \sum_{n,m=0}(-1)^{m+1}(n+1)r^{(m)}_n \right) \,
    .
\end{equation}
Here $t_2$ is a number of the Lagrangian equations, $r^{(m)}_n$ is
the number of the gauge symmetry generators of the total order $n$
and reducibility order $m$. In the irreducible case,
($m=0$) this brings the well known relation for the degrees of
freedom for the second order Lagrangian equations \cite{HTZ}:
\begin{equation}\label{DoFL2o1}
   \frac{\mathcal{N}}{2}=  t_2 \, - \, \sum_{n=0}(n+1)r_n \, .
\end{equation}
$\mathcal{N}/2$ has the meaning of number of physical polarizations.
%where $r_n$ is a number of the gauge transformations of the
% order $n$. Notice that the relations (\ref{DoFL2}),
%(\ref{DoFL2o1}) provide the correct count for the physical degrees
%of freedom thus far the Lagrangian equations are involutive from the
%outset. If the Lagrangian equations are not involutive, the general
%formula (\ref{DF-A}), being applied to the involutive closure of the
%equations, has to provide a correct count for the physical degrees
%of freedom which can be different from (\ref{DoFL2}),
%(\ref{DoFL2o1}).

\end{document}